\documentclass[
  10pt,
  twocolumn,
  showpacs,
  preprintnumbers,
  amsmath,
  amssymb,
  aps,
  prd,
  nofootinbib,
  superscriptaddress,
  longbibliography,
  floatfix
]{revtex4-1}

\usepackage[english]{babel}
\usepackage[T1]{fontenc}

\usepackage{graphicx}
\usepackage{epstopdf}

\usepackage{amsmath}
\usepackage{amssymb}
\usepackage{amsfonts}
\usepackage{mathrsfs}
\usepackage{bm}

\usepackage{dcolumn}
\usepackage{multirow}
\usepackage{enumerate}

\usepackage{xcolor}
\usepackage{wasysym}

\usepackage[caption=false]{subfig}

\usepackage{hyperref}
\hypersetup{
  colorlinks=true,
  linkcolor=blue,
  citecolor=green,
  urlcolor=blue
}

\begin{document}

\title{Optical Signatures of Black Holes Surrounded by a Generalized Cloud of Strings}


\author{Marcos V. de S. Silva}
\email{marcos.sousa@uva.es}
\affiliation{Department of Theoretical Physics, Atomic and Optics, Campus Miguel Delibes, \\ University of Valladolid UVA, Paseo Bel\'en, 7, 47011 - Valladolid, Spain}

\author{Ednaldo L. B. Junior} \email{ednaldobarrosjr@gmail.com}
\affiliation{Faculdade de F\'{i}sica, Universidade Federal do Pará, Campus Universitário de Tucuruí, CEP: 68464-000, Tucuruí, Pará, Brazil}
\affiliation{Programa de P\'{o}s-Gradua\c{c}\~{a}o em F\'{i}sica, Universidade Federal do Sul e Sudeste do Par\'{a}, 68500-000, Marab\'{a}, Par\'{a}, Brazil}

\author{Carlos F. S. Pereira} 
	\email{carlosfisica32@gmail.com}
	\affiliation{Departamento de Física, Universidade Federal do Espírito Santo, Avenida Fernando Ferrari, 514, Goiabeiras, 29060-900, Vit\'oria, Espir\'ito Santo, Brazil.}

\author{Diego S\'aez-Chill\'on G\'omez}
\email{diego.saez@uva.es} 
\affiliation{Department of Theoretical Physics, Atomic and Optics, and Laboratory for Disruptive Interdisciplinary Science (LaDIS), Campus Miguel Delibes, \\ University of Valladolid UVA, Paseo Bel\'en, 7,
47011 - Valladolid, Spain}
\affiliation{Departamento de F\'isica, Universidade Federal do Cear\'a, Caixa Postal 6030, Campus do Pici, 60455-760 Fortaleza, Cear\'a, Brazil.}
     

\begin{abstract}
In this work, we investigate the optical signatures of black holes surrounded by a generalized cloud of strings, described by the Letelier--Alencar spacetime. We first analyze null geodesics, showing that the string parameters can either increase or decrease the photon sphere radius and the critical impact parameter relative to their Schwarzschild values. Since the spacetime is not asymptotically flat, the critical impact parameter does not directly coincide with the physical shadow radius. We use this distinction to derive constraints on the model parameters from the shadow radius bounds of Sgr A* and M87*. In the weak-field regime, we calculate the light deflection angle using both the Gauss--Bonnet method and a perturbative expansion of the null geodesic equations, obtaining consistent results that contain a contribution from the asymptotically conical geometry in addition to the local gravitational deflection. We also examine the Shapiro time delay and identify a distance dependent contribution associated with the non Minkowskian asymptotic background. Finally, we study the optical appearance of geometrically and optically thin accretion disks and construct celestial sphere images through backward ray tracing. The transfer functions and lensing images show that the generalized cloud of strings modifies the null geodesic mapping in a nonuniform manner rather than producing a simple rescaling of the Schwarzschild image. In particular, configurations with a smaller shadow may still produce an outward displacement of the main Einstein ring image, demonstrating that different optical observables probe distinct aspects of photon propagation in this geometry.
\end{abstract}

\date{\today}

\maketitle

\section{Introduction}

General Relativity predicts the existence of black holes as a natural consequence of its field equations. These objects represent the extreme regime of gravitational interaction, where the nonlinear effects of gravity become dominant. Within the classical framework, their formation is associated with the gravitational collapse of sufficiently massive stars or other compact objects, for which the forces responsible for maintaining hydrostatic equilibrium are unable to counterbalance gravitational collapse \cite{Penrose:1964wq}. The first exact analytical description of a black hole was obtained by Karl Schwarzschild, who derived the vacuum solution corresponding to the exterior gravitational field of a static, spherically symmetric mass distribution \cite{Schwarzschild:1916uq}. Since then, several generalizations have been developed to describe more realistic physical scenarios, including the presence of electric charge, a cosmological constant, and angular momentum \cite{Wald:1984rg,dInverno:1992gxs}. Among these extensions, the Kerr solution describes a rotating vacuum black hole with axial symmetry, whereas the Kerr--Newman solution simultaneously incorporates the effects of rotation and electric charge. In the Kerr geometry, the fundamental properties of the spacetime are completely determined by the mass and angular momentum parameters. In contrast, the Kerr--Newman solution introduces an additional parameter associated with the electric charge, thus enlarging the parameter space of the solution \cite{Kerr:1963ud,Newman:1965my}. Owing to their theoretical significance and observational relevance, these solutions have served as the foundation for the study of compact astrophysical objects and a wide range of phenomena associated with gravitational fields, both in the weak- and strong-field regimes.

Over the past decades, the study of the optical appearance of black holes has become one of the most promising observational probes for investigating gravity in the strong-field regime. Long before imaging the immediate vicinity of the event horizon became feasible, theoretical studies had already demonstrated that the propagation of photons near these compact objects is strongly influenced by the curvature of spacetime. Depending on their initial conditions, light rays may either escape to infinity, be captured by the black hole, or follow critical trajectories associated with unstable null orbits. The boundary separating these distinct classes of trajectories gives rise to a dark region in the observed image, whose morphology encodes valuable information about the underlying spacetime geometry and the nature of the gravitational field \cite{Synge:1966okc,Luminet:1979nyg,Chandrasekhar:1985kt,Bardeen:1973tla}.

Observations of light emissions from the vicinity of M87* and Sgr A*, conducted by the Event Horizon Telescope (EHT), have transformed this theoretical prediction into a powerful observational tool \cite{EventHorizonTelescope:2019dse,EventHorizonTelescope:2022wkp}. This is because both the angular size of the black hole shadow and the structure of the surrounding bright ring are directly determined by the photon capture region of the central compact object, thereby providing a unique opportunity to probe the geometry of spacetime in the immediate vicinity of the event horizon. The acquisition of the first image of a black hole shadow by the EHT renewed interest in the study of the optical properties of compact objects and gravitational lensing phenomena, further highlighting their importance for observational astrophysics. This landmark observation underscored the need for increasingly accurate theoretical models capable of describing the propagation of light through regions of extreme spacetime curvature, characteristic of the vicinity of compact objects \cite{EventHorizonTelescope:2019ggy,EventHorizonTelescope:2019dse,EventHorizonTelescope:2022wkp,EventHorizonTelescope:2022xqj}.

Significant advances in the description of the strong-field regime were achieved through the work of Virbhadra and Ellis \cite{Virbhadra:1999nm,Perlick:2007nt}. They developed a formalism to describe gravitational lensing produced by compact objects in asymptotically flat spacetimes, demonstrating that sufficiently strong gravitational fields can deflect light rays to such an extent that multiple images of the same background source are formed. Moreover, this formalism established the foundation for the systematic investigation of strong-field gravitational lensing, becoming a benchmark for numerous studies devoted to the optical properties of black holes and other compact objects. Other pioneering work relevant to the development of gravitational lensing theory and its applications in a variety of gravitational scenarios can be found in Refs. \cite{Virbhadra:2002ju,Claudel:2000yi,Virbhadra:1998dy,Virbhadra:2008ws,Virbhadra:2007kw,Bozza:2002zj,Bozza:2001xd,Tsukamoto:2016jzh}. Furthermore, in the context of investigating the optical properties and light deflection in black hole spacetimes, numerous recent studies have explored the effects of various extensions of General Relativity and novel physical scenarios, including quantum gravity \cite{Soares:2025hpy,Junior:2023xgl,Kumar:2025mpb,Liu:2024oas,Dong:2024alq}, noncommutative geometries \cite{AraujoFilho:2025huk,Filho:2024zxx,Feng:2024jeq,AraujoFilho:2025rvn}, dark matter halos \cite{Ahmed:2025nfi,AraujoFilho:2026ajq,Molla:2025yoh,Ali:2025ney,Qiao:2024ehj,Macedo:2024qky}, Lorentz symmetry violation \cite{Junior:2024vdk,Pereira:2026lyt,Pantig:2025eda,Jha:2024xtr,Atamurotov:2022wsr,Gao:2024ejs,Ovgun:2025ctx,Nascimento:2026rho}, and topological defects \cite{Soares:2024rhp,Soares:2023uup,Lan:2025hlv,Lu:2025mcm}, among others. Related studies have also investigated accretion disks around regular black holes \cite{Akbarieh:2023kjv,Khoshrangbaf:2025bwg} and the optical properties of exotic compact objects such as wormholes and black bounces \cite{Alencar:2026gcl}.

Astrophysical black holes are not expected to exist in complete isolation, but rather within environments containing accretion flows, electromagnetic fields, dark matter distributions, and possibly other matter fields. When the gravitational backreaction of the surrounding material is taken into account, the resulting spacetime generally differs from the corresponding vacuum black hole geometry. Such configurations are commonly referred to as ``dirty black holes.'' Depending on the properties and distribution of the external matter, the backreaction may modify the horizon structure, thermodynamic quantities, orbital dynamics, and the response of the system to perturbations, thereby leaving signatures in both electromagnetic and gravitational wave observations
\cite{Abramowicz:2011xu,Visser:1992qh,Barausse:2014pra,Cardoso:2021wlq,Lui:2026nxy}.

The optical properties of a dirty black hole are determined by the complete backreacted geometry rather than by the vacuum metric of the central object alone. Since null geodesics are sensitive to the radial distribution and compactness of the surrounding matter, environmental effects may shift the circular photon orbits, modify the critical impact parameter and the photon capture region, and alter the resulting shadow and gravitational lensing pattern \cite{Fonseca:2025ehf}. Matter concentrated sufficiently close to the black hole can produce appreciable deviations from the corresponding vacuum geometry and, in highly compact configurations, may even generate additional light rings and more intricate lensing structures
\cite{Konoplya:2019sns,Xavier:2023exm}.

An interesting example of a dirty black hole is obtained when the surrounding matter is described by a cloud of strings. In this case, a macroscopic cloud of one-dimensional objects is represented as a continuous anisotropic matter distribution. Since strings carry energy density and tension along a preferred spatial direction, the associated effective stress-energy tensor generally exhibits different radial and transverse pressures. The first exact static and spherically symmetric black hole solution sourced by such a distribution was obtained by Letelier \cite{Letelier:1979ej}. In the original Letelier configuration, the radial pressure is equal in magnitude and opposite in sign to the energy density, whereas the transverse pressures vanish. The cloud of strings introduces an additional constant contribution to the metric function, causing the resulting spacetime to differ from the Schwarzschild geometry even at large radial distances.

Since Letelier's work, black holes surrounded by a cloud of strings have been investigated in several gravitational settings. The presence of the string distribution has been shown to modify the horizon structure and thermodynamic properties of black holes, including their phase behavior and stability \cite{MoraisGraca:2018ofn,Nascimento:2023tgw}. The effects have also been explored in the motion of massive and massless particles, where they modify the effective potential, circular orbits, and gravitational lensing \cite{Nascimento:2023tgw,Kumar:2025mpb,Vishvakarma:2024icz}. In addition, their influence on quasinormal modes, black hole shadows, and optical observables has been analyzed in different geometries \cite{Singh:2022ycn,Kumar:2025mpb,Vishvakarma:2023csw}. These studies demonstrate that a cloud of strings can affect both the causal structure of the spacetime and the propagation of particles and radiation around the black hole.

More recently, Alencar et al. generalized Letelier's construction by allowing an additional magnetic-like component of the string bivector, in addition to the electric-like component considered in the original solution \cite{Alencar:2025zyl}. The resulting anisotropic source is characterized by an effective coupling $g_s$ and a length scale $l_s$. In contrast to the original Letelier model, the generalized stress-energy tensor possesses nonvanishing tangential pressures, while the additional length scale modifies the radial distribution of the source and the causal structure of the spacetime. The original Letelier solution is recovered in the limit $l_s\rightarrow 0$, with $g_s^2$ identified with the standard cloud of strings parameter.

The generalized cloud of strings framework has subsequently been investigated in several contexts, including wormhole constructions, geodesic motion, tidal effects, regularized black hole solutions, and gravitational lensing
\cite{Kumar:2025mpb,Ahmed:2025nfi,Alencar:2026dsq,Silva:2025hwl,Muniz:2025ugk,Zhong:2026kqy}. In this work, we extend the investigation of this geometry through a comprehensive study of its optical properties, including the analysis of null geodesics, the modifications of the black hole shadow, light deflection in the weak-field regime, the time delay, and other related optical observables.

This work is organized as follows: In Sec.~\ref{SEC:spacetime}, we review the Letelier spacetime and the Letelier--Alencar generalization. In Sec.~\ref{SEC:NG}, through the study of null geodesics, we analyze the behavior of the photon-sphere radius and the critical impact parameter. In Sec.~\ref{SEC:SR}, we compute the black hole shadow radius and compare it with EHT data for M87* and Sgr A*. In Sec.~\ref{SEC:LD}, we use the weak-field approximation to calculate the light deflection and the Shapiro time delay. In Sec.~\ref{SEC:OA}, we analyze the optical appearance of a black hole surrounded by a geometrically and optically thin accretion disk and subsequently construct celestial sphere images through backward ray tracing. Finally, our conclusions are presented in Sec.~\ref{Sec:Conclusion}.


\section{Letelier--Alencar spacetime}\label{SEC:spacetime}
To investigate the optical properties of black holes surrounded by a cloud of strings, we first need to establish the background geometry. We consider a static, spherically symmetric spacetime defined by the general line element:
\begin{equation}
ds^2 = -f(r)dt^2 + \frac{dr^2}{f(r)} + r^2(d\theta^2+\sin^2\theta d\phi^2).\label{line}
\end{equation}

While several black hole configurations coupled with cloud of strings exist in the literature, our focus is on the original formulation by Letelier and its recent generalization by Alencar and collaborators. The standard Letelier model describes a black hole immersed in an anisotropic distribution of strings \cite{Letelier:1979ej}. The dynamics of this background are governed by the Nambu-Goto action,
\begin{equation}
    S_{NG} = \int\mathcal{M} \sqrt{-\gamma}\, d\lambda^0 d\lambda^1,
\end{equation}with $\mathcal{M}$ representing a dimensionless string constant, $\lambda^A$ (for $A = 0, 1$) being the world-sheet coordinates, and $\gamma$ denoting the determinant of the induced metric $\gamma_{AB}$.

Through a macroscopic averaging of this string distribution, an effective stress-energy tensor emerges. By defining an antisymmetric bivector $\Sigma^{\mu\nu}$ and ensuring stress-energy conservation, Letelier derived an exact analytical solution for the metric function:
\begin{equation}
f(r) = 1 - \frac{2M}{r} - \alpha,
\end{equation}
where $M$ is a mass parameter and the parameter $\alpha$ (restricted to $0 < \alpha < 1$) encodes the presence of the cloud of strings. This parameter effectively modifies the gravitational scenario, shifting the event horizon to $r_h = 2M / (1 - \alpha)$. The matter source exhibits a purely radial pressure profile \cite{Rodrigues:2022rfj}. This type of solution also arises in other contexts, such as spacetimes involving Kalb--Ramond fields, nonlinear electrodynamics, and global monopoles \cite{Barriola:1989hx,Yang:2023wtu,Alencar:2026dsq}.

A fundamental assumption in Letelier's original derivation was the strict negativity of $\gamma$, which restricted the bivector to only have a non-vanishing $\Sigma^{01}$ component \cite{Letelier:1979ej}. Recently, Alencar et al. demonstrated that the condition $\gamma < 0$ can be maintained even if angular components are introduced, specifically allowing $\Sigma_{23} \neq 0$ \cite{Alencar:2025zyl}. In this generalized framework, the bivector components are parameterized by a string length scale $l_s$ and an effective coupling constant $g_s^2$, which dictates the gravitational strength of the cloud at tree level. The resulting stress-energy tensor components take the form \cite{Alencar:2025zyl}:
\begin{equation}T^t{}_t = T^r{}_r = - \frac{g_s^2}{r^4}\sqrt{l_s^4+r^4}, \quad T^\theta{}_\theta = T^\phi{}_\phi = \frac{g_s^2l_s^4}{r^4\sqrt{l_s^4+r^4}}.
\end{equation}
When evaluating the limit where $l_s \to 0$ and $g_s^2 \to \alpha$, this model reduces to the standard Letelier case.

Solving the Einstein field equations with this generalized source yields a modified metric function:
\begin{equation}
f(r) = 1 - \frac{2M}{r} + \frac{g_s^2 l_s^2}{r^2} \, {}_2F_{1}\left(-\frac{1}{2}, -\frac{1}{4}, \frac{3}{4}, -\frac{r^4}{l_s^4}\right),
\end{equation}
where ${}_2F_{1}$ denotes the hypergeometric function.

Although the horizon roots for $f(r)=0$ cannot be isolated analytically, the causal structure presents rich features. For coupling values of $g_s < 1$, the spacetime can have both an event horizon and a Cauchy horizon, sharing similarities with the Reissner--Nordström geometry. Depending on $l_s$, the inner and outer horizons can degenerate to an extremal limit. However, if the effects associated with a large $l_s$ overcome gravitational attraction, the event horizon vanishes, exposing a naked singularity.

\section{Null Geodesics and Photon Orbits}\label{SEC:NG}
The trajectories followed by a photon can be described through the geodesics of a massless particle.

To obtain the photon trajectories, we consider the Lagrangian associated with the line element~\eqref{line}, given by
\begin{equation}
    \mathcal{L}=\frac{1}{2}\left(-f(r)\dot{t}^2 + \frac{\dot{r}^2}{f(r)} + r^2\dot{\theta}^2+r^2\sin^2\theta \dot{\phi}^2\right)=\kappa,\label{Lag}
\end{equation}
where the dot represents the derivative with respect to an affine parameter $\lambda$ and $\kappa=-1/2$ for timelike geodesics or $\kappa=0$ for null geodesics. Since our spacetime is spherically symmetric, we can, without loss of generality, restrict the photon motion to the equatorial plane, $\theta=\pi/2$. From the form of Eq.~\eqref{Lag}, we see that the Lagrangian does not depend explicitly on the coordinates $t$ and $\phi$. Therefore, we have two conserved quantities associated with the Lagrangian, given by
\begin{equation}
    E=f\dot{t},\quad \mbox{and} \quad L= r^2\dot{\phi}.\label{const}
\end{equation}
The conserved quantities $E$ and $L$ are associated with the energy and angular momentum of the orbits, respectively.

\begin{figure*}[htb]
    \centering
    \includegraphics[width=.5\linewidth]{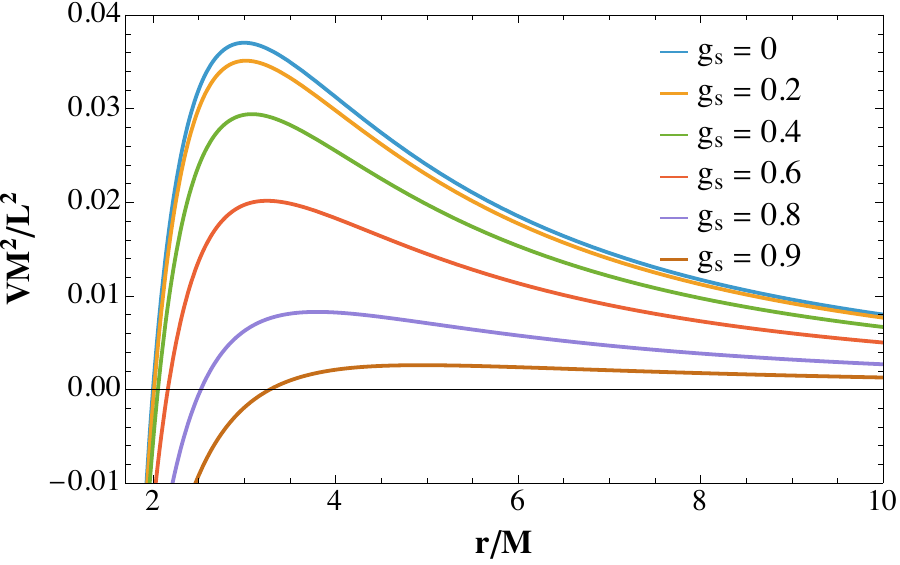}\hspace{-0.1cm}
    \includegraphics[width=.5\linewidth]{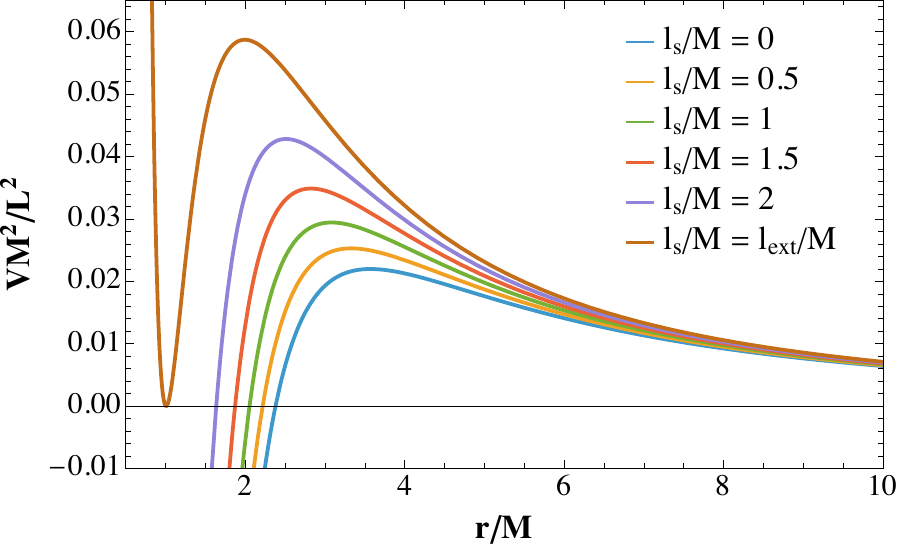}
    \caption{Behavior of the effective potential for photons as a function of the radial coordinate. In the left panel, we fix $l_s/M=1$ and vary the value of $g_s$. In the right panel, we fix $g_s=0.4$ and vary the value of $l_s$.}
    \label{fig:pot}
\end{figure*}

Since photons are massless particles, we can use the relations obtained in Eq.~\eqref{const} and rewrite Eq.~\eqref{Lag} for null geodesics. In this way, we obtain
\begin{equation}\label{motion}
    E^2=\dot{r}^2+V, \quad \mbox{with} \quad V=\frac{fL^2}{r^2},
\end{equation}
where $V$ is the effective potential of the system. Equation~\eqref{motion} describes the dynamics of photons moving under the influence of the effective potential $V$. In Fig.~\ref{fig:pot}, we show the behavior of the effective potential as the parameters associated with the cloud of strings are varied. The effective potential vanishes at the event horizon. As the value of $g_s$ increases, the horizon radius also increases, causing the potential to change sign at larger distances from the center of the black hole. In addition, the maximum of the effective potential decreases with increasing $g_s$. As a result, particles require less energy to overcome the potential barrier and be absorbed by the black hole. When varying $l_s$, we find that larger values of $l_s$ lead to a higher effective potential, implying that particles need more energy to overcome the potential barrier. Furthermore, increasing $l_s$ shifts the event horizon toward smaller radial coordinates, causing the effective potential to vanish closer to the center of the black hole. Expanding the effective potential for $r\to \infty$ we find
\begin{equation}
    V(r)\approx-\frac{\left(g_s^2-1\right) L^2}{r^2}+O\left(\frac{1}{r^3}\right).
\end{equation}
This behavior suggests that the effects of the cloud of strings are not localized near the black hole, but instead extend to spatial infinity.

The maximum point shown in Fig.~\ref{fig:pot} corresponds to an unstable circular orbit. In other words, particles approaching the black hole with an energy equal to the maximum of the effective potential can move along a circular orbit around the black hole. When a particle follows a closed circular orbit, the conditions $\left.\dot{r}\right|_{r=r_c}=0$ and $\left.\ddot{r}\right|_{r=r_c}=0$ must be satisfied, where $r_c$ is the radius of the circular orbit, which imply $\left.V\right|_{r=r_c}=E^{2}$ and $\left.V'\right|_{r=r_c}=0$. From these conditions, we can obtain the radius of the circular orbit and the critical impact parameter through the expressions
\begin{equation}
    2f(r_c)-r_cf'(r_c)=0, \quad b_c=\frac{L_c}{E_c}=\frac{r_c}{\sqrt{f(r_c)}}.\label{LR_condition}
\end{equation}
The impact parameter is defined as $b=L/E$, while the critical impact parameter corresponds to the value associated with unstable circular orbits. 

For the Schwarzschild case, i.e. $g_s\to0$, the relation \eqref{LR_condition} gives us the well-known result for the radius of the photon sphere, $r_c=3M$. For the Letelier case, $l_s\to 0$, we obtain $r_c=3M/(1-g_s^2)$. However, in the Letelier--Alencar case, Eq.~\eqref{LR_condition} cannot be solved analytically, although it can be solved numerically.

\begin{figure*}[htb]
    \centering
    \includegraphics[width=0.5\linewidth]{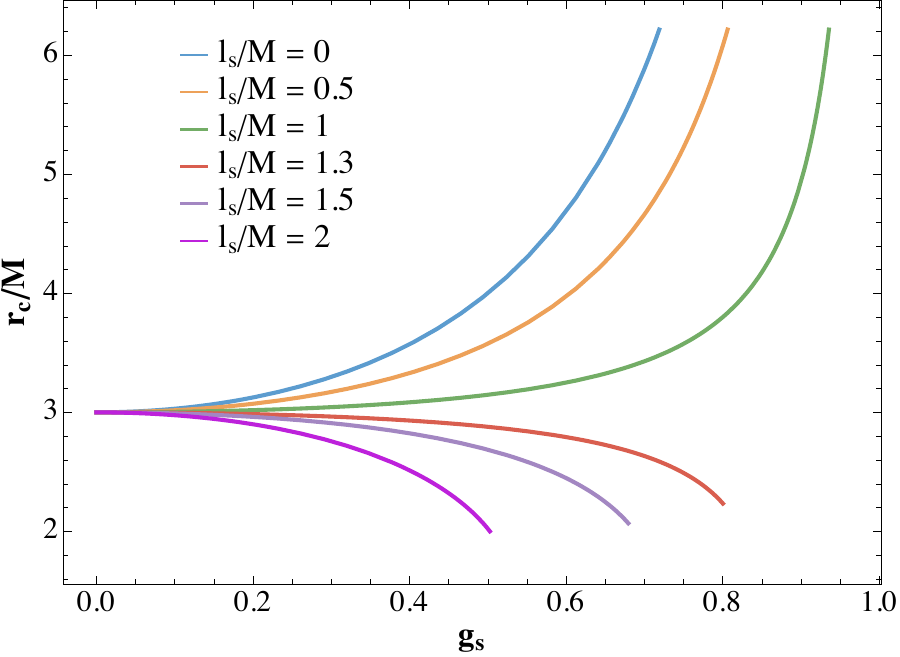}\hspace{-0.1cm}
    \includegraphics[width=0.5\linewidth]{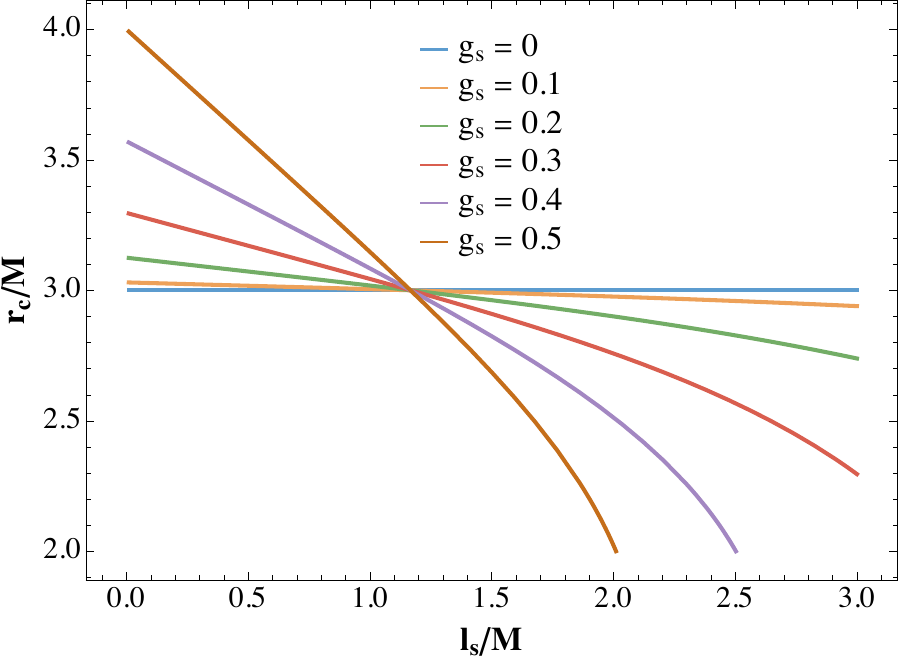}
    \caption{Behavior of the photon sphere radius for different values of the cloud of strings parameters. In the left panel, we fix different values of $l_s$ and analyze the behavior of the curves as $g_s$ is varied. In the right panel, we fix different values of $g_s$ and examine how the curves behave as $l_s$ is varied.}
    \label{fig:photon_sphere}
\end{figure*}

In Fig.~\ref{fig:photon_sphere}, we show the behavior of the photon sphere radius as the cloud of strings parameters are varied. As $g_s\to0$, the Schwarzschild result is recovered, namely $r_c=3M$. The behavior of the radius of the photon sphere as a function of $g_s$ also depends on the value of $l_s$. For small values of $l_s$, the radius of the light ring increases as $g_s$ increases. However, this growth becomes progressively weaker for larger values of $l_s$. If $l_s$ is sufficiently large, the radius of the unstable orbit instead decreases as $g_s$ increases. The lower panel confirms that $l_s$ plays the role of reducing $r_c$. For a fixed value of $g_s$, larger values of $l_s$ lead to smaller orbital radii. Moreover, increasing $g_s$ produces larger photon sphere radii when $l_s$ is small, whereas for sufficiently large $l_s$ the opposite behavior is observed, resulting in smaller orbital radii.

\begin{figure*}[htb]
    \centering
    \includegraphics[width=0.5\linewidth]{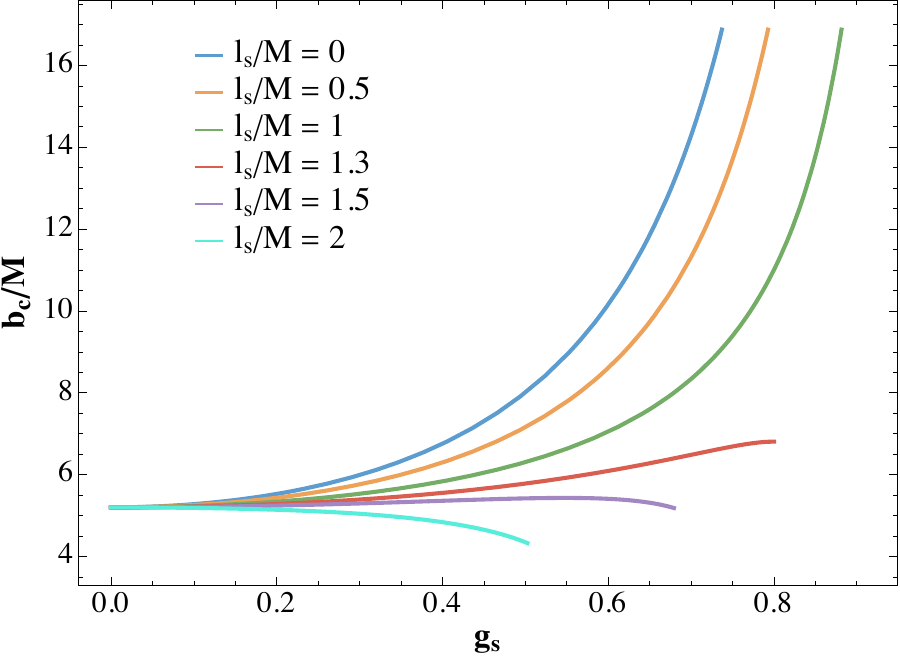}\hspace{-0.1cm}
    \includegraphics[width=0.49\linewidth]{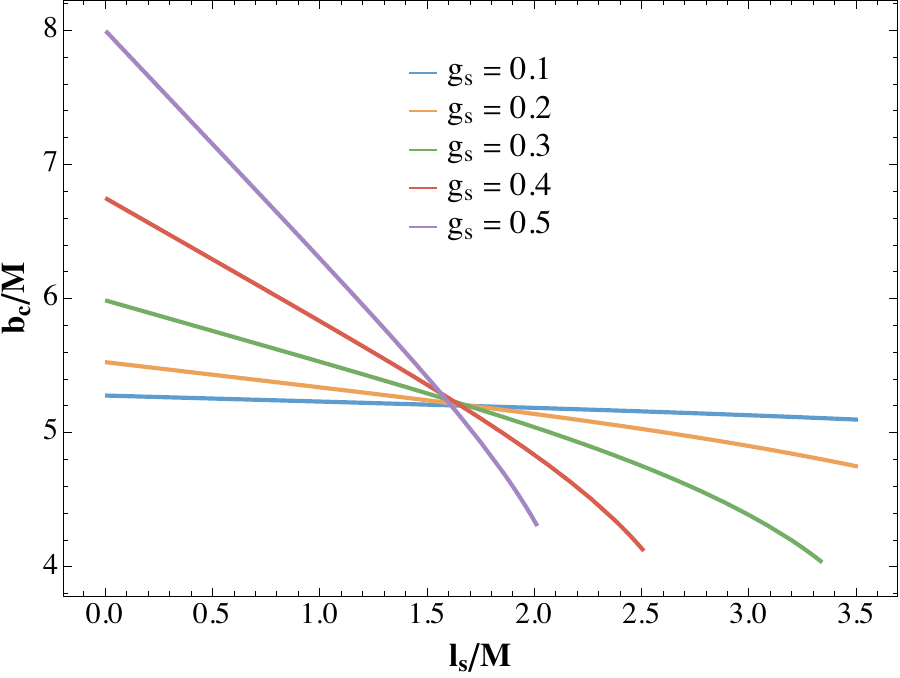}
    \caption{Behavior of the critical impact parameter for different values of the cloud of strings parameters. In the upper panel, we fix different values of $l_s$ and analyze the behavior of the curves as $g_s$ is varied. In the lower panel, we fix different values of $g_s$ and examine how the curves behave as $l_s$ is varied.}
    \label{fig:bc}
\end{figure*}

In addition to the analysis of $r_c$, it is also important to examine the behavior of $b_c$, since for spherically symmetric spacetimes the critical impact parameter essentially represents the shadow radius of the black hole for distant observers. Moreover, it is directly related to the geometric absorption cross section through $\sigma_{\rm geo}=\pi b_c^2$.

Figure~\ref{fig:bc} shows that the behavior of $b_c$ closely follows that of $r_c$. The Schwarzschild result is recovered as $g_s\to0$, and $b_c$ increases with increasing $g_s$ provided that $l_s$ is sufficiently small. The parameter $l_s$ tends to suppress this growth and, for sufficiently large values, causes $b_c$ to decrease as $g_s$ increases. The lower panel clearly shows that $b_c$ decreases as $l_s$ increases, and this reduction becomes more pronounced for larger values of $g_s$.

\begin{figure*}[!htb]
    \centering
    \includegraphics[width=0.5\linewidth]{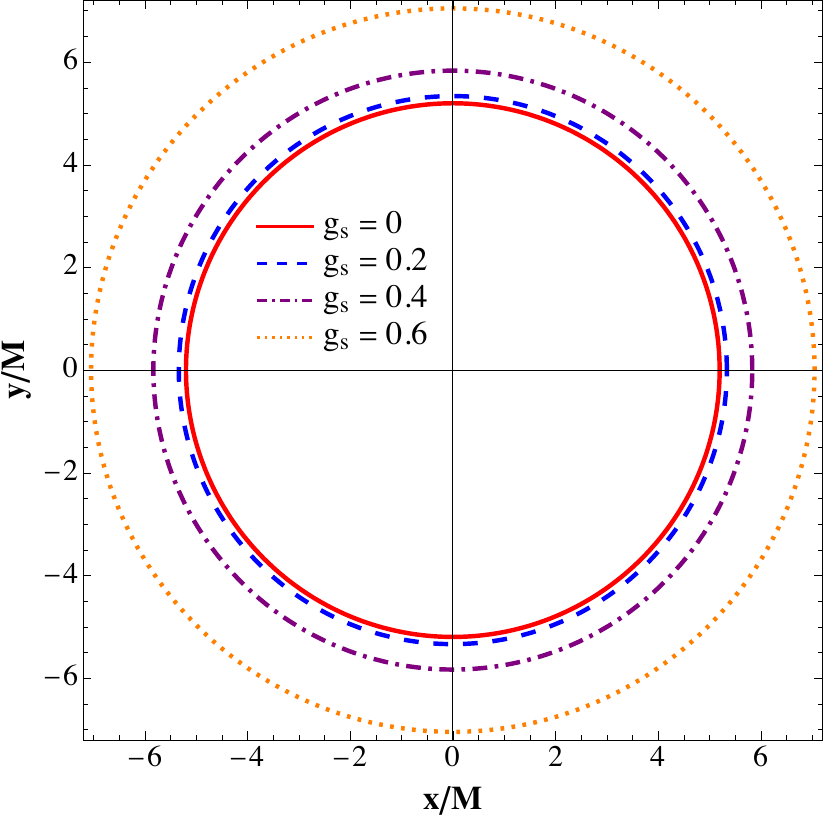}\hspace{-0.1cm}
    \includegraphics[width=0.5\linewidth]{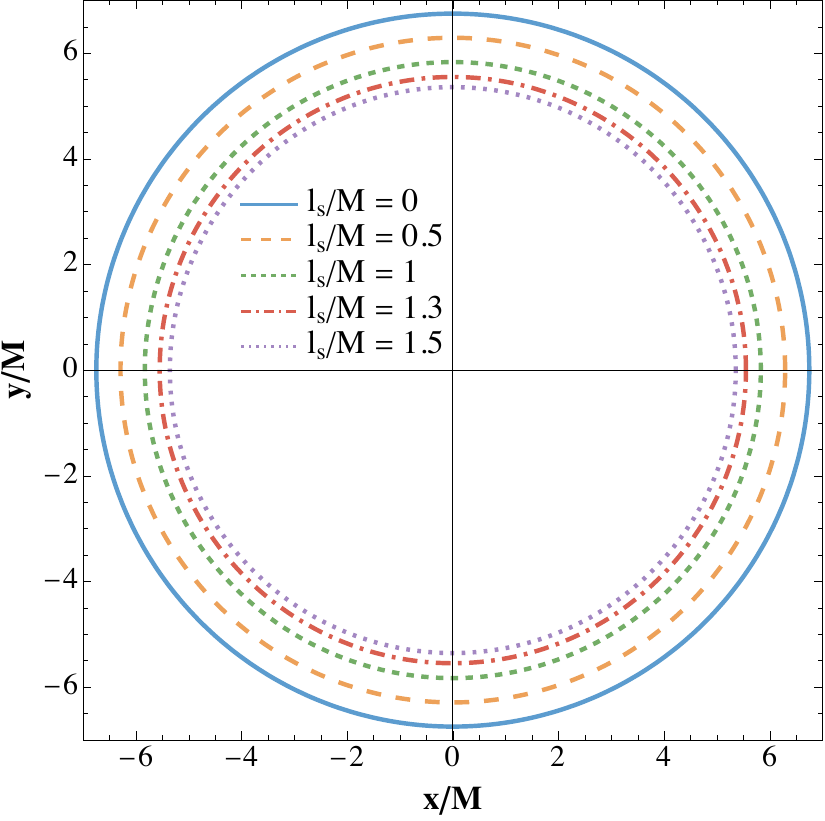}
    \caption{Behavior of the shape formed by the critical impact parameter for the Letelier--Alencar spacetime for different values of the model parameters. In the upper panel, we fix $l_s/M=1$ and vary the value of $g_s$, whereas in the lower panel, we fix $g_s=0.4$ and vary the values of $l_s$.}
    \label{fig:shape}
\end{figure*}

In Fig.~\ref{fig:shape}, we present the corresponding shape formed by the critical impact parameter for different values of the model parameters. A more detailed discussion of the shadow radius, including a comparison with the EHT observations, will be presented later.

To conclude this section, we investigate the behavior of photon trajectories in this geometry. To this end, we rewrite Eq.~\eqref{motion} using the coordinate transformation $r=1/u$. This yields
\begin{equation}\label{geo1}
    \left(\frac{du}{d\phi}\right)^2=\frac{1}{b^2}-u^2f(u).
\end{equation}
We can also work with the equivalent second-order differential equation, which is given by
\begin{equation}
    \frac{d^2u}{d\phi^2}=-\frac{u^2}{2}\frac{df(u)}{du}-uf(u).\label{geo2}
\end{equation}

\begin{figure*}[!htb]
    \centering
    \includegraphics[width=0.45\linewidth]{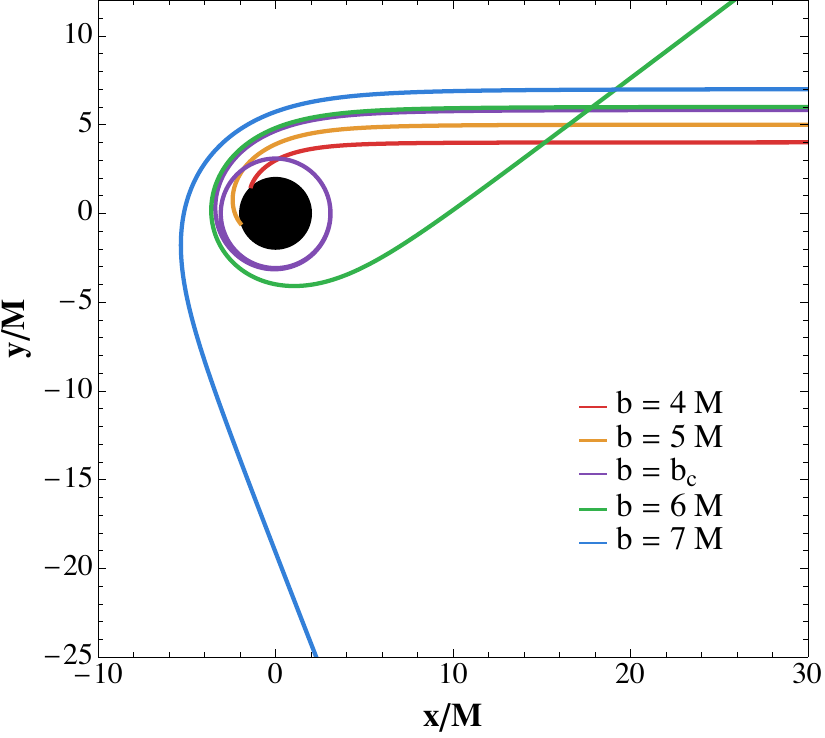}
    \includegraphics[width=0.478\linewidth]{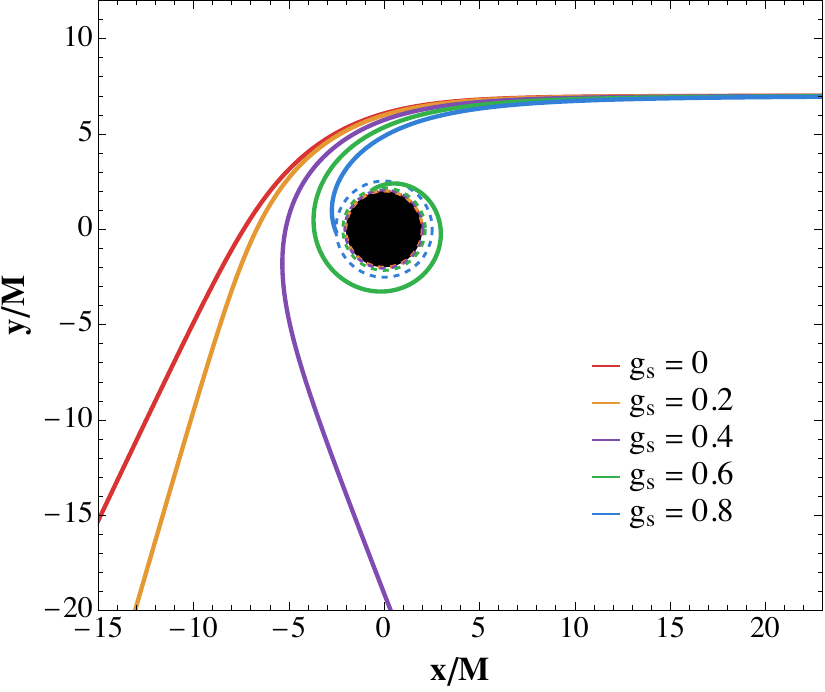}
    \includegraphics[width=0.45\linewidth]{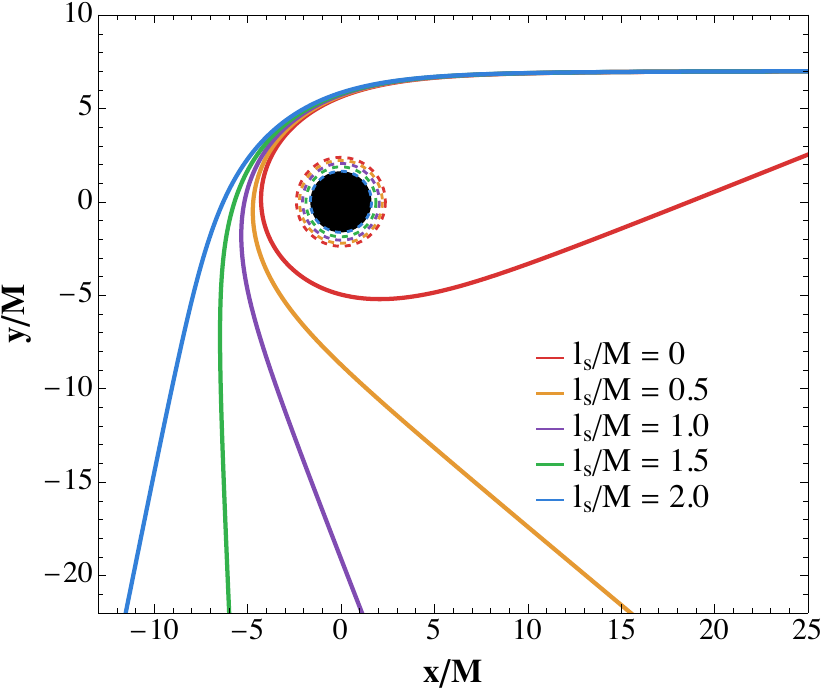}\caption{Photon trajectories in the Letelier--Alencar spacetime for different values of the cloud of strings parameters. In the upper panel, we fix $g_s=0.4$ and $l_s/M=1$ while varying the impact parameter. In the middle panel, we fix $b=7M$ and $l_s/M=1$ and vary $g_s$. In the lower panel, we fix $b=7M$ and $g_s=0.4$ and vary the values of $l_s$. In the middle and lower panels, the dashed circles indicate the event horizon radius for each set of parameter values. The dark disk represents the area enclosed by the smallest event horizon.}
    \label{fig:geodesics}
\end{figure*}

With the appropriate boundary conditions, Eq.~\eqref{geo2} can be solved numerically, allowing us to obtain the photon trajectories in the Letelier--Alencar spacetime. The corresponding results are shown in Fig.~\ref{fig:geodesics}. As is well known from the literature, particles with $b<b_c$ are absorbed by the black hole, particles with $b>b_c$ are scattered by the black hole, and particles with $b=b_c$ follow an unstable circular orbit around the black hole. When we vary $g_s$ while keeping $l_s$ and $b$ fixed, we find that the scattering experienced by the photons is modified, leading to an increase in the deflection angle. Eventually, particles that would otherwise be scattered become absorbed by the black hole. This behavior is consistent with the results obtained from the effective potential, since the potential barrier decreases as $g_s$ increases. On the other hand, increasing $l_s$ reduces the deflection angle, which is also in agreement with the effective-potential analysis. In this case, the potential barrier becomes higher as $l_s$ increases, implying that particles require even greater energy to overcome the barrier and be absorbed by the black hole.

\section{Shadow Radius}\label{SEC:SR}
As discussed previously, for spherically symmetric spacetimes the critical impact parameter can be interpreted as the radius of the black hole shadow for observers located sufficiently far from the black hole \cite{Vagnozzi:2022moj}. We have also shown that this quantity changes when the cloud of strings parameters are varied. More precisely, however, the shadow radius depends on the position of the observer relative to the black hole.

The black hole shadow radius has been widely used in conjunction with EHT observations as a tool to constrain the parameters of modified theories of gravity and the additional parameters that arise in alternative black hole models \cite{Vagnozzi:2022moj}. Therefore, we can also employ this method to place constraints on the parameters associated with the cloud of strings.

To obtain the shadow radius, we follow the procedure proposed by Perlick and Tsupko \cite{Perlick:2021aok,Vagnozzi:2022moj}. For this purpose, we first define the function\footnote{For more general spherically symmetric spacetimes of the form $g_{00}=-A(r)$, $g_{11}=B(r)$, $g_{22}=D(r)$, and $g_{33}=D(r)\sin^{2}\theta$, the function $h(r)$ is defined as
$h(r)=\sqrt{\frac{D}{A}}$.}
\begin{equation}
    h(r)=\sqrt{\frac{r^2}{f(r)}}.\label{hr}
\end{equation}
This function can be related to the impact parameter through the relation $h(r_0)=b$, where $r_0$ denotes the turning point of the photon trajectory. It can also be related to the critical impact parameter by taking the limit $r_0 \to r_c$.

The angular radius of the black hole shadow, $\alpha_{sh}$, is related to the function $h$ by the relation \cite{Perlick:2021aok}
\begin{equation}
    \sin^2 \alpha_{sh}=\frac{h^2(r_c)}{h^2(r_O)},
\end{equation}
where $r_{O}$ represents the location of the observer. Using Eq.~\eqref{hr}, we can rewrite the angular radius as
\begin{equation}
   \sin^2 \alpha_{sh}=\frac{b_c^2}{h^2(r_O)}=\frac{r_c^2f(r_O)}{f(r_c)r_O^2}. 
\end{equation}
For physically relevant situations, in which the observer is located very far from the black hole, we may take the small-angle limit. Therefore, the radius of the black hole shadow can be written as
\begin{equation}
    r_{sh}=r_c\sqrt{\frac{f(r_O)}{f(r_c)}}.
\end{equation}

Determining how the shadow radius behaves is not sufficient to constrain the cloud of strings parameters. We also need to impose astrophysical constraints on this radius. Due to observational uncertainties, the measured value of the black hole shadow radius will not be exactly equal to its theoretical prediction. Nevertheless, it is possible to quantify how large the deviation between these values can be considered acceptable. For the Schwarzschild case, the percentage difference between them can be quantified through the relation
\begin{equation}
    \delta=\frac{r_{sh,ob}}{3\sqrt{3}M}-1,
\end{equation}
where $r_{sh,ob}$ is the observed shadow radius. Using the estimates for $\delta$, it is possible to constrain the allowed values of the shadow radius of observed black holes, such as M87* and Sgr A*. For these cases, the ranges obtained for the validity region of the shadow radius are
\begin{equation}
    4.55 \lesssim \frac{r_{s}}{M} \lesssim 5.22 \quad (1\sigma), 
    \quad
    4.21 \lesssim \frac{r_{s}}{M} \lesssim 5.56 \quad (2\sigma),
\end{equation}
for Sgr A* and
\begin{equation}
    4.26 \lesssim \frac{r_{s}}{M} \lesssim 6.03 \quad (1\sigma), 
    \quad
    3.38 \lesssim \frac{r_{s}}{M} \lesssim 6.91 \quad (2\sigma),
\end{equation}
for M87*. These results are obtained from EHT, Keck, and VLTI data \cite{EventHorizonTelescope:2019dse,Do:2019txf,GRAVITY:2020gka}. These observational inputs will be used to derive constraints on the cloud of strings parameters.

\begin{figure*}[!htb]
    \centering
    \includegraphics[width=0.5\linewidth]{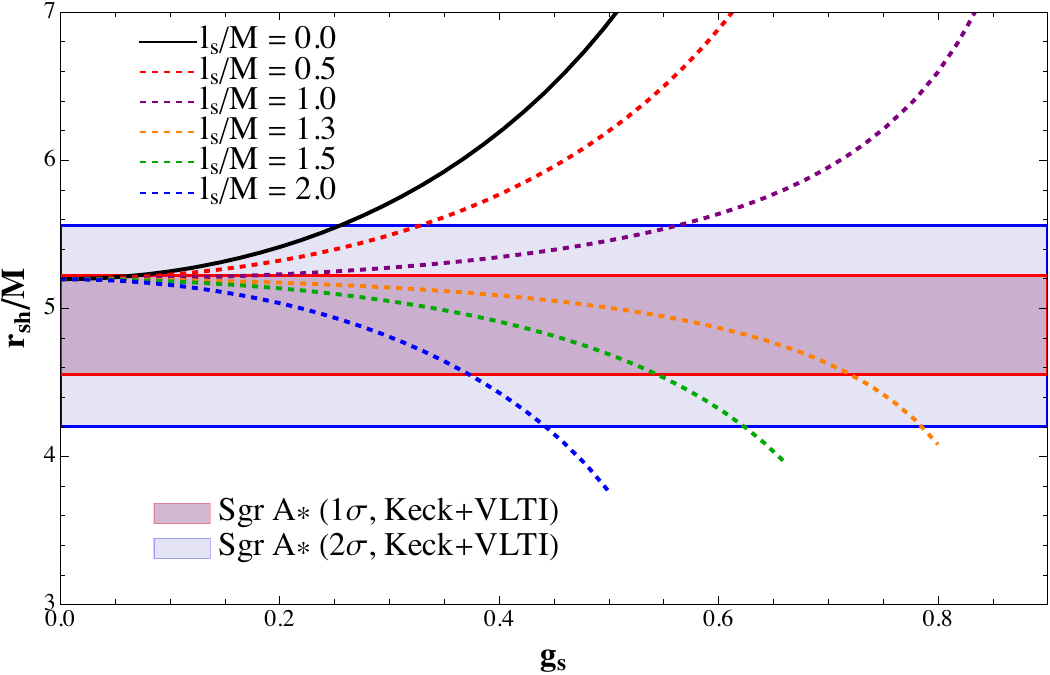}\hspace{-0.1cm}
    \includegraphics[width=0.5\linewidth]{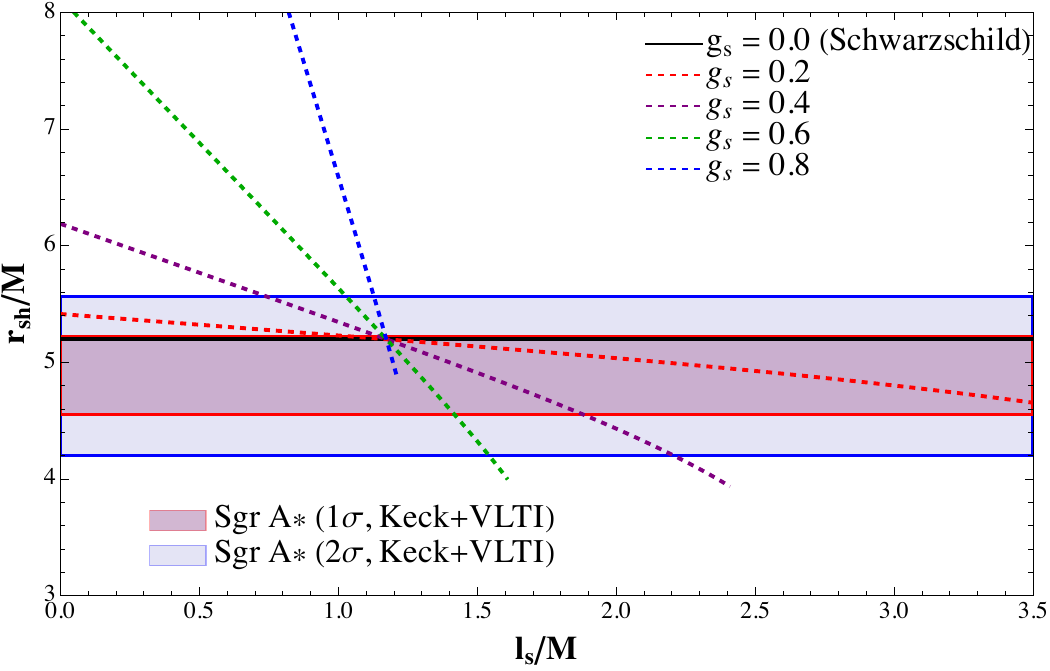}
    \caption{Shadow radius of the black hole for the Letelier--Alencar spacetime. The red and blue regions represent the intervals consistent with the shadow radius of Sgr A$^\ast$ obtained from the Keck+VLTI data at the $1\sigma$ and $2\sigma$ confidence levels, respectively. In the upper panel, we fix different values of $l_s$ and vary the parameter $g_s$. In the lower panel, we fix different values of $g_s$ and vary the parameter $l_s$.
}
    \label{fig:sh_sagA}
\end{figure*}

In Fig. \ref{fig:sh_sagA}, we show the behavior of the shadow radius using the Keck+VLTI data for the shadow radius of Sgr A$^\ast$ for different values of the cloud of strings parameter. First, we fix different values of $g_s$ and vary $l_s$. We observe that, for small values of $g_s$, the shadow radius remains compatible with the observational bounds. For $g_s=0.2$, the shadow radius remains within the $2\sigma$ confidence interval for the entire range of $l_s$ considered. The allowed range of $l_s$ becomes more restricted as $g_s$ increases. For $g_s=0.4$, $l_s$ is constrained to the interval $0.74 \lesssim l_s/M \lesssim 2.20$. If we further increase $g_s$, the allowed range of $l_s$ becomes even narrower. For the case in which $g_s$ is varied, we observe that all curves tend to the Schwarzschild limit as $g_s \to 0$. As $g_s$ increases, the curves tend to move away from the confidence region allowed by the observational results. For intermediate values of $l_s$, such as $l_s/M=1.4$, the shadow radius remains within the $1\sigma$ and $2\sigma$ confidence intervals over a large range of $g_s$, up to approximately $g_s=0.78$. On the other hand, if $l_s$ is either very small, such as $l_s/M=0$, or very large, such as $l_s/M=2$, the corresponding curves tend to leave the confidence region more rapidly.

\begin{figure*}[!htb]
    \centering
    \includegraphics[width=0.5\linewidth]{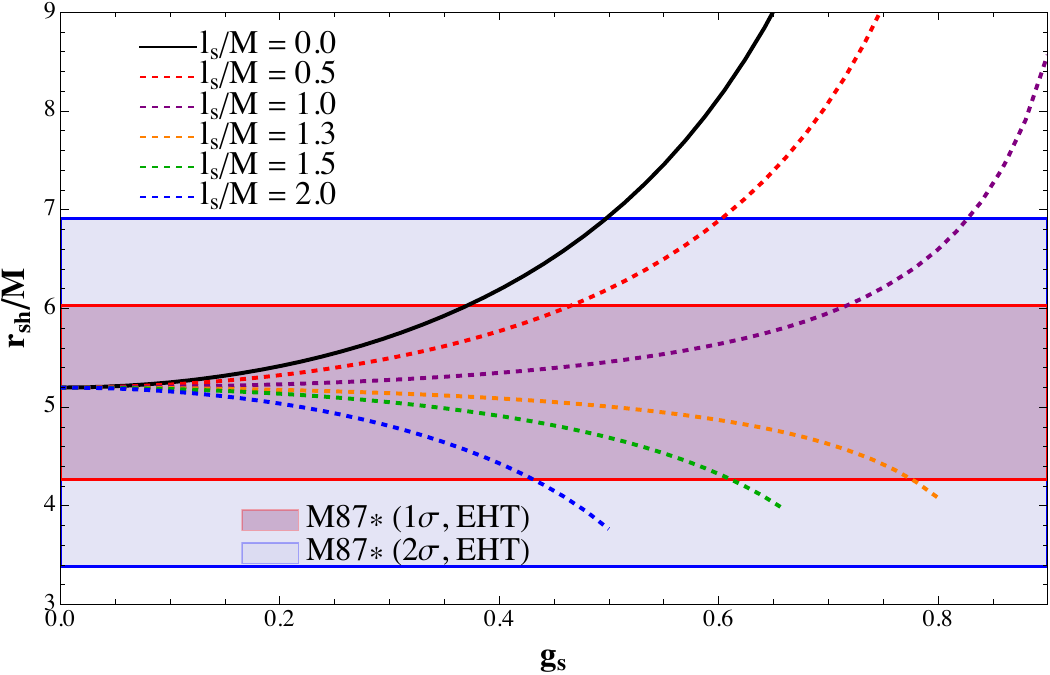}\hspace{-0.1cm}
    \includegraphics[width=0.5\linewidth]{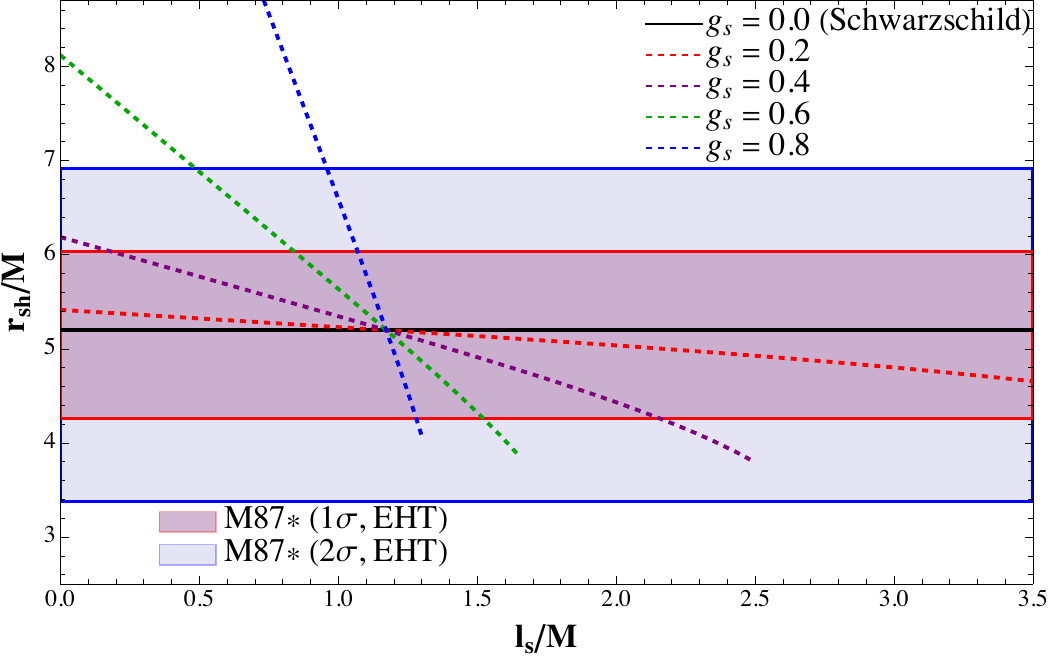}
    \caption{Shadow radius of the black hole for the Letelier-Alencar spacetime. The red and blue regions represent the intervals consistent with the shadow radius of M87$^\ast$ obtained from the EHT collaboration data at the $1\sigma$ and $2\sigma$ confidence levels, respectively. In the upper panel, we fix different values of $l_s$ and vary the parameter $g_s$. In the lower panel, we fix different values of $g_s$ and vary the parameter $l_s$.
}
    \label{fig:sh_M87}
\end{figure*}

In Fig. \ref{fig:sh_M87}, we repeat the analysis of the shadow radius, but now using the observational data for the black hole M87$^\ast$ obtained by the EHT Collaboration. The results are similar to those found for Sgr A$^\ast$. The main difference arises from the confidence intervals associated with the observational data. These intervals are larger for M87$^\ast$, which leads to weaker constraints on the model parameters. This can be seen, for instance, in the case where we set $g_s=0.2$ and vary $l_s$. In this example, for small values of $l_s$, the curve starts within the $2\sigma$ region and then enters the $1\sigma$ region as $l_s$ increases for the Sgr A$^\ast$ data. In contrast, for the M87$^\ast$ data, the same curve remains within the $1\sigma$ region over the entire range of $l_s$ considered. Therefore, the Sgr A$^\ast$ data provide a more stringent observational filter for constraining the parameters of our black hole model.

Another point that needs to be discussed is the difference between the results for the shadow radius and for the critical impact parameter. As mentioned previously, for spherically symmetric and asymptotically flat spacetimes, the critical impact parameter essentially represents the shadow radius. However, by comparing our results in Figs.~\ref{fig:bc}, \ref{fig:sh_sagA}, and \ref{fig:sh_M87}, we see that they do not necessarily coincide. This happens because, in asymptotically flat spacetimes, one has $f(r)\to 1$. In our model, however, this is not the case, since $f(r)\to 1-g_s^2$. Therefore, this factor enters the expression for the black hole shadow radius, making it differ from the result obtained for the critical impact parameter.


\section{Weak-field light deflection and Shapiro time delay}\label{SEC:LD}

In this section, we investigate light propagation in spacetimes containing a cloud of strings, focusing on two characteristic weak-field effects: the deflection of light and the Shapiro time delay. Although light propagation can be studied in both the strong- and weak-field regimes, here we restrict our analysis to the weak-field regime, since strong-field gravitational lensing in this geometry has already been extensively investigated in Ref.~\cite{Kumar:2025mpb}.

The weak deflection angle is calculated using two different approaches. The first is based on the Gauss--Bonnet method \cite{Gibbons:2008rj}, which relates the deflection angle to the geometrical properties of the optical spacetime. The second consists of a direct weak-field expansion of the equations of motion governing null geodesics. Comparing the results obtained from these two methods provides a consistency check of the weak-field approximation. We then analyze the Shapiro time delay by evaluating the additional propagation time experienced by light rays due to the spacetime geometry.

\subsection{Weak deflection angle with the Gauss--Bonnet approach}
To construct the Gauss--Bonnet method, we consider the line element given by Eq.~\eqref{line}. In the equatorial plane, this line element can be written as
\begin{equation}
ds^2 = -f(r)dt^2 + \frac{dr^2}{f(r)} + r^2d\phi^2.
\end{equation}
Since for photons we have $ds^2=0$, the equation can be rewritten as
\begin{equation}
d\sigma^2=dt^2= g_{\mu\nu}^{opt}dx^\mu dx^\nu=\frac{dr^2}{f^2(r)} + \frac{r^2}{f(r)}d\phi^2,
\end{equation}
where $d\sigma^2$ denotes the line element associated with the optical metric $g_{\mu\nu}^{opt}$.

Associated with this line element, we can compute the area element of the optical metric, which is given by
\begin{equation}
    dS=\sqrt{g^{opt}}drd\phi=\frac{r}{f(r)^{3/2}}drd\phi,
\end{equation}
where $g^{opt}$ is the determinant of the metric $g_{\mu\nu}^{opt}$.

Another important quantity in the Gauss--Bonnet method for light deflection is the Gaussian curvature, $\mathcal{K}$, which for the optical metric is given by
\begin{equation}
    \mathcal{K}=\frac{R}{2}=\frac{1}{2}ff''-\frac{1}{4}f'^2.
\end{equation}

At this point, an important feature of the Letelier--Alencar solution must be emphasized. If
\begin{equation}
f(r\to\infty) = f_\infty \neq 1,
\end{equation}
then the asymptotic optical metric is not Euclidean. In the limit $r\to \infty$, one has
\begin{equation}
d\sigma^2 \simeq \frac{dr^2}{f_\infty^2}+\frac{r^2}{f_\infty}d\phi^2 .
\end{equation}
In the Letelier--Alencar case, the asymptotic value is given by
\begin{equation}
f_\infty = 1-g_s^2 .
\end{equation}
This form shows that the asymptotic optical geometry is not Euclidean, but rather conical. Therefore, the usual straight line approximation $r=b/\sin\phi$, valid for asymptotically flat spacetimes, must be replaced by the corresponding zeroth order trajectory in the conical background.

Defining
\begin{equation}
\rho=\frac{r}{f_\infty},
\end{equation}
we obtain
\begin{equation}
d\sigma^2 \simeq d\rho^2+f_\infty \rho^2 d\phi^2 .
\end{equation}
If we write
\begin{equation}
\beta=\sqrt{f_\infty},
\end{equation}
then
\begin{equation}
d\sigma^2 \simeq d\rho^2+\rho^2 d\varphi^2,
\end{equation}
where $d\varphi^2=\beta^2d\phi^2$.  Through the angular variable $\varphi$
the asymptotic optical metric becomes locally Euclidean.
In this auxiliary Euclidean plane, the undeflected light ray is described by a straight line. Written in terms of the original radial coordinate ($r$) and angular coordinate ($\phi$), this trajectory is
$
r(\phi)=\frac{b\beta}{\sin(\beta\phi)}$. The corresponding angular interval is $0\leq \phi \leq \frac{\pi}{\beta}$. For $\beta=1$, the asymptotically flat result $r=b/\sin\phi$ is immediately recovered.

Another important point is that, in the weak-field regime, we can approximate the metric function $f$ and the Gaussian curvature $\mathcal{K}$ as
\begin{equation}
    f(r) \approx \left(1-g_s^2\right)-\frac{2 M}{r}+\frac{2g_s^2 l_s \Gamma \left(\frac{3}{4}\right)^2}{\sqrt{\pi }r}+O\left(\frac{1}{r^3}\right), \label{ffraco}
\end{equation}
\begin{equation}
    \mathcal{K}\approx\frac{2 \left(g_s^2-1\right) \left(M-\frac{g_s^2 l_s \Gamma \left(\frac{3}{4}\right)^2}{\sqrt{\pi }}\right)}{r^3}+O\left(\frac{1}{r^4}\right).
\end{equation}

The total weak deflection angle can then be written as \cite{Gibbons:2008rj}
\begin{equation}
\hat{\alpha}_{GB} =\pi\left(\frac{1}{\beta}-1\right)-\frac{1}{\beta}
\int_{0}^{\pi/\beta}
\int_{r(\phi)}^{\infty}
\mathcal{K}dS .
\end{equation}
Substituting $r(\phi)$ and integrating,
we obtain
\begin{equation}
\begin{split}
\hat{\alpha}_{GB}
=& \pi\left(\frac{1}{\beta}-1\right)
-\frac{4 g_s^2 l_s \Gamma \left(\frac{3}{4}\right)^2-4 \sqrt{\pi } M}{b \sqrt{\pi }\beta ^4},\\
=& \pi\left(\frac{1}{\sqrt{1-g_s^2}}-1\right)+\frac{4 M}{b(1-g_s^2)^{2}}-\frac{4 g_s^2 l_s \Gamma \left(\frac{3}{4}\right)^2}{b \sqrt{\pi }(1-g_s^2)^{2}}.\label{GB}
\end{split}
\end{equation}
The first term is the contribution coming from the conical structure of the optical geometry at infinity, while the second term is the local contribution due to the Gaussian curvature of the optical metric.

In the Schwarzschild limit $g_s\to 0$, one has $\beta\to 1$, and the standard result
$\hat{\alpha}_{GB}\to \frac{4M}{b}$
is recovered. In the Letelier limit, namely $l_s\to 0$, we obtain
\begin{equation}
\hat{\alpha}_{GB}= \pi\left(\frac{1}{\sqrt{1-g_s^2}}-1\right)+\frac{4 M}{b(1-g_s^2)^{3/2}}.
\end{equation}
which is the expected result for the Letelier case, or equivalently for the case with a topological charge. The inclusion of this additional term in the calculation of the light deflection angle for non-asymptotically flat spacetimes was first proposed by Kimet Jusufi et al. in Ref. \cite{Jusufi:2017lsl}. Since then, this formulation has been widely adopted and further extended in numerous subsequent studies, establishing itself as an important framework for investigating the optical properties of non-asymptotically flat geometries \cite{Pereira:2026lyt,Pereira:2026ffn,Jusufi:2017xnr,Jusufi:2017hed,Jusufi:2017mav,Ono:2018jrv,Soares:2023err,Furtado:2020puz}.

\subsection{Expansion for light deflection in the weak-field limit}
In the Gauss--Bonnet method, we essentially use geometrical properties of the spacetime, such as the Gaussian curvature and the optical metric. The geodesic equations are not used directly. Therefore, a useful way to confirm the weak-field deflection result is to derive the same expression through an alternative method, which in this case consists of a direct expansion of the results obtained from the geodesic equations.

To this end, we recover Eq.~\eqref{geo1} and integrate it. In the case of light deflection, the contributions before and after the turning point are equal. Therefore, we can consider the integration from infinity to the turning point of the photon. With this, we obtain
\begin{equation}
    \Delta \phi=2\int_{0}^{u_0}\left(\frac{1}{b^2}-u^2f(u)\right)^{-1/2}du.
\end{equation}
Performing the appropriate weak-field expansions, keeping only first-order terms in $l_s$ and $M$, and relating the impact parameter to the turning point through $b=1/u_0\sqrt{f(u_0)}$, we find
\begin{equation}
\delta\phi
= \pi\left(\frac{1}{\sqrt{1-g_s^2}}-1\right)+\frac{4 M}{b(1-g_s^2)^{2}}
-\frac{4 g_s^2 l_s \Gamma \left(\frac{3}{4}\right)^2}{b \sqrt{\pi }(1-g_s^2)^{2}}.\label{weak}
\end{equation}
In the expression above, the angular variation is computed as $\delta\phi=\Delta\phi-\pi$, where we subtract from the total angular variation the angle that the particle would have described in the absence of curvature due to the mass or to the topological defect. Comparing Eq.~\eqref{GB} with Eq.~\eqref{weak}, we see that, at least to first order in the expansions in $l_s/b$ and $M/b$, the weak-field deflection obtained from the direct expansion of the equation of motion derived from the geodesic equations agrees with the weak-field result obtained through the Gauss--Bonnet method.

\subsection{Shapiro Time Delay }\label{shap}

To evaluate the Shapiro time delay, we consider the propagation of a light ray between two observers located at radial coordinates $r_1$ and $r_2$. Assuming a superior conjunction geometry, with $r_1, r_2\gg d$, where $d$ denotes the photon distance of closest approach to the black hole, the total round-trip propagation time is given by
\begin{equation}
T=2\left[t(r_1,d)+t(r_2,d)\right].
\label{totalt}
\end{equation}
where the total propagation time from $d$ to a point $r_i$ is given by \cite{Junior:2023nku}
\begin{equation}
    t(r_i,d) = \frac{1}{b} \int_{d}^{r_i}   \frac{1}{f(r)} \Biggl[ \frac{1}{b^2} - \frac{f(r)}{r^2} \Biggr]^{-1/2}dr\,.
    \label{eq:tintegrado}
\end{equation}
where
\begin{equation}
    b = \sqrt{\frac{d^2}{f(d)}}\,,
    \label{eq:bemd}
\end{equation}
is determined at the distance of closest approach of the light ray to the black hole. 
For the metric function \eqref{ffraco}, the photon travel time can be evaluated directly, which in turn allows the corresponding Shapiro time delay to be computed, 
\begin{eqnarray}
&&t(r)=
\sqrt{r^{2}-d^{2}}
\left[
1+
g_s^2+\frac{\left(M-\frac{g_s^2l_s}{\sqrt{\pi}}\Gamma\left(\frac{3}{4}\right)^{2}\right)}
{r+d}
\right]\nonumber\\
&&+2\left(M-\frac{g_s^2l_s}{\sqrt{\pi}}\Gamma\left(\frac{3}{4}\right)^{2}\right)
\operatorname{arctanh}\left(
\frac{r}{\sqrt{r^{2}-d^{2}}}
\right).
\label{eq:t_r}
\end{eqnarray}
This delay is defined as the difference between the actual propagation time and the reference travel time in the absence of the gravitational field. Since the spacetime described by the string-inspired solution is not asymptotically Minkowski, owing to the nonvanishing contribution of the parameter $g_s$ at spatial infinity, two complementary definitions of the Shapiro time delay can be considered. The first,
\begin{equation}
\Delta T \equiv T-T_0\,, \label{delT1}
\end{equation}
which $T_0$ is the corresponding travel time in flat spacetime, measures the total propagation delay with respect to flat spacetime, thereby accounting for both the global modification of the background geometry induced by the string sector and the gravitational field of the black hole.  Using Eqs. \eqref{totalt}, \eqref{eq:t_r} and \eqref{delT1}, we obtain
\begin{equation}  
\begin{split}
&\Delta T
\simeq4M
\left[1+\ln\!\left(\frac{4r_1 r_2}{d^2}\right)
\right]\\
&+2g_s^2
\left(
r_1+r_2\right)-\frac{4g_s^2l_s}{\sqrt{\pi}}\Gamma\!\left(\frac{3}{4}\right)^2\left[1+
\ln\!\left(\frac{4r_1 r_2}{d^2}\right)
\right]\,.
\label{eq:DeltaTcor_approx}
\end{split}
\end{equation}
The standard Schwarzschild result is immediately recovered in the limit $g_s \rightarrow 0$, whereas the cloud of string contribution remains nonvanishing in the limit $l_s \rightarrow 0$. The resulting time delay naturally contains two qualitatively distinct contributions: a logarithmic term associated with the localized gravitational field of the black hole, which generalizes the standard Shapiro delay, and a distance-dependent term proportional to $g_s^2(r_1+r_2)$, arising from the non-Minkowskian asymptotic structure generated by the cloud of string. Similar distance-dependent terms have also been reported in other non asymptotically flat spacetimes \cite{Cardenas:2021eri}.  

To quantify the impact of the string cloud, it is convenient to compare the present prediction with the Schwarzschild result through the relative deviation
\begin{eqnarray}
\delta T&=&\frac{\Delta T-\Delta T_{\rm Sch}}{\Delta T_{\rm Sch}}\nonumber\\
&=&\frac{g_s^2}{M}\left[\frac{(r_1+r_2)}{2
\left[
1+
\ln\!\left(\frac{4r_1 r_2}{d^2}\right)
\right]}-\frac{l_s}{\sqrt{\pi}}\Gamma\!\left(\frac{3}{4}\right)^2\right]\,.\label{deltT}
\end{eqnarray}
Note that if $\delta T=0$, the Schwarzschild result is recovered. For $\delta T>0$, the cloud of string increases the Shapiro time delay, whereas for $\delta T<0$, it reduces the delay. The first term in \eqref{deltT} represents the cumulative delay due to the propagation of light through the background generated by the string cloud, while the second, negative term decreases the coefficient of the logarithmic contribution, effectively reducing the gravitational mass. Furthermore, if $r_1$ and $r_2$ are parameterized in terms of $M$, then for supermassive black holes the second term becomes practically negligible, and the relative correction approaches a constant value when the distances are expressed in units of the gravitational radius. This indicates that the leading relative correction is governed primarily by the cloud of string parameters rather than by the mass of the central object.

The numerical results presented in Table~\ref{tab:ShapiroNum} are fully consistent with the analytical expression derived for the Shapiro time delay. First, the time delay for Schwarzschild to Sgr A$^\ast$, with $M=4.297\times10^6 M_\odot$, is  $\Delta T_{\rm Sch}= 938.517$s and for M87$^\ast$, with $M=6.5\times10^9\,M_\odot$, is $\Delta T_{\rm Sch}= 1.41968\times10^6$s \footnote{Throughout the calculations, geometric units with $G=c=1$ are adopted. The solar mass is expressed as a length through $M_\odot=G M_\odot^{\rm SI}/c^2$, and the black hole mass is then given in units of $M_\odot$. The radial distances are expressed in the same units, so that the calculated time delays have dimensions of length and are converted to physical times in seconds by dividing by $c$.
}. The total delay and the additional contribution due to the cloud of strings both increase with the black hole mass. Since the source and observer distances are chosen as fixed multiples of the gravitational radius, i.e., $r_1/ M$ and $r_2/ M$, the dominant correction, proportional to $g_s^2(r_1+r_2)$, scales linearly with the black hole mass. Consequently, the additional time delay becomes significantly larger for supermassive black holes, indicating that these systems constitute the most promising astrophysical laboratories for probing the effects of a surrounding cloud of strings through light propagation experiments.

The Table~\ref{tab:ShapiroNum} also reveals that the correction is strongly governed by the parameter $g_s$. Keeping $l_s$ fixed, the additional delay increases approximately as $g_s^2$, in agreement with the analytical prediction. In contrast, varying $l_s$ over the interval considered produces changes only in the eleventh decimal place for Sgr A$^\ast$ and in the ninth decimal place for M87$^\ast$. This behavior reflects the fact that the $l_s$ dependent contribution enters only through the subleading logarithmic term, whereas the dominant correction grows linearly with the propagation distance. Therefore, for realistic astrophysical configurations, the Shapiro time delay is essentially insensitive to $l_s$, making $g_s$ the parameter that primarily controls the observable deviations from the Schwarzschild prediction.

\begin{table}[!t]
\centering

\begin{tabular}{cccc}
\hline
Black hole & $(g_s,l_s/M)$ &
$\Delta T_{\rm }$ &
$\Delta T_{\rm }-\Delta T_{\rm Sch}$ \\
\hline

\multirow{8}{*}{Sgr A$^\ast$}

& $(0.2,1.5)$ & 14484.011 & 13545.494 \\
& $(0.4,1.5)$ & 55120.500 & 54182.000 \\
& $(0.6,1.5)$ & 122848.000 & 121909.000 \\
& $(0.8,1.5)$ & 217666.000 & 216728.000 \\

\cline{2-4}

& $(0.3,0.5)$ & 31415.878282 & 30477.361041 \\
& $(0.3,1.3)$ & 31415.878282 & 30477.361041 \\
& $(0.3,1.5)$ & 31415.878282 & 30477.361041 \\
& $(0.3,2.5)$ & 31415.878282 & 30477.361041 \\

\hline

\multirow{8}{*}{M87$^\ast$}

& $(0.2,1.5)$ & $2.19097\times10^7$ & $2.04900\times10^7$ \\
& $(0.4,1.5)$ & $8.33798\times10^7$ & $8.19602\times10^7$ \\
& $(0.6,1.5)$ & $1.85830\times10^8$ & $1.84410\times10^8$ \\
& $(0.8,1.5)$ & $3.29260\times10^8$ & $3.27841\times10^8$ \\

\cline{2-4}

& $(0.3,0.5)$ & $4.75223\times10^7$ & $4.61026\times10^7$ \\
& $(0.3,1.3)$ & $4.75223\times10^7$ & $4.61026\times10^7$ \\
& $(0.3,1.5)$ & $4.75223\times10^7$ & $4.61026\times10^7$ \\
& $(0.3,2.5)$ & $4.75223\times10^7$ & $4.61026\times10^7$ \\

\hline
\end{tabular}
\caption{Numerical estimates of the Shapiro time delay for Sgr A$^\ast$ and M87$^\ast$. The calculations were performed assuming $r_1=5000M$, $r_2=3000M$, and $d=50M$. All times are given in seconds.}
\label{tab:ShapiroNum}
\end{table}

\section{Optical Appearance}\label{SEC:OA}
In the previous sections, we analyzed the main optical properties associated with null geodesics in the Letelier--Alencar spacetime. These results characterize several aspects of photon propagation in this geometry and provide the basis for studying the observable appearance of the black hole.

In this section, we investigate the optical signatures of the generalized cloud of strings from two complementary perspectives. First, we consider a black hole surrounded by a geometrically and optically thin accretion disk and analyze the observed intensity distribution produced by different emission profiles. We then study gravitational lensing through backward ray tracing of a celestial sphere, which provides a direct visualization of how the spacetime geometry maps the observer's local sky into distant light sources.

\subsection{Optically thin accretion disk}
The image of a black hole is not determined only by the existence of a photon sphere or by the size of the critical impact parameter. Instead, it depends on the interplay between the background geometry and the astrophysical properties of the emitting matter surrounding the compact object \cite{Guerrero:2021ues,Gralla:2019xty,Guerrero:2022qkh,Guerrero:2022msp,Olmo:2021piq}. In this sense, the accretion disk provides the source of radiation, while the spacetime geometry determines how the emitted photons propagate before reaching a distant observer. The combination of these effects gives rise to the observed intensity distribution on the observer's screen.

We consider a geometrically and optically thin accretion disk located on the equatorial plane of the black hole. The disk is assumed to emit radiation isotopically with a radial intensity profile $I^{em}(r)$. Since the disk is optically thin, photons may cross the equatorial plane several times before reaching the observer. Each crossing gives a possible contribution to the observed image. The first intersection is associated with the direct emission, while subsequent intersections correspond to lensed and photon ring contributions.

Since the detailed physics of accretion flows is highly model-dependent, we adopt a phenomenological description for the emitted intensity. A convenient choice is provided by the Standard Unbound distribution \cite{Gralla:2020srx}
\begin{equation}
I^{em}_{SU}(r,\mu,\sigma,\gamma)=\frac{
e^{-\frac{1}{2}
\left[\gamma+\operatorname{arcsinh}\left(\frac{r-\mu}{\sigma}\right)\right]^2}}{\sqrt{(r-\mu)^2+\sigma^2}}.\label{perfis}
\end{equation}
The parameter $\mu$ controls the location of the emission peak, $\sigma$ controls the width of the emitting region, and $\gamma$ determines the asymmetry of the profile. This family of profiles is useful because it allows us to model different physical situations while keeping the disk emission analytically simple. This profile has been widely used to investigate the optical appearance of black holes and other compact objects \cite{Alencar:2026qeb,Nojiri:2026auy,Olmo:2025ctf,Olmo:2023lil,Aimar:2025jxr,Macedo:2025ipc,Macedo:2024qky,Rosa:2023qcv}.

For the Letelier--Alencar spacetime, there are three natural radial scales that can be used to define representative emission profiles: the event horizon radius $r_h$, the photon sphere radius $r_c$, and the innermost stable circular orbit (ISCO)\footnote{In this work, we do not investigate the motion of massive particles in the Letelier--Alencar geometry. However, this analysis is carried out in detail in Ref.~\cite{Silva:2025hwl}.
} $r_{I}$. The first scale is associated with emission concentrated close to the black hole horizon. The second one emphasizes the region where strong lensing effects become relevant. The third one represents a disk whose emission is dominated by matter orbiting near the inner edge of a standard thin accretion disk. We recall that the horizon position is obtained by solving $f(r_h)=0$. The photon orbit radius is determined from the condition \eqref{LR_condition}, while the ISCO radius can be obtained from the relation
\begin{equation}
2r_{I}f(r_{I})f''(r_{I})-4rf'(r_{I})^2+6f(r_{I})f'(r_{I})=0.
\label{eq:cond-isco-en}
\end{equation}

Accordingly, we consider three representative luminosity profiles. The first profile has its emission peak located close to the event horizon,
$\mu_1\simeq r_h$. This profile models a disk whose emission extends deep into the gravitational potential well. Since the emitted photons originate very close to the black hole, this profile is expected to be strongly affected by gravitational redshift and by the existence of the shadow region. The second profile is peaked near the photon sphere,
$\mu_2\simeq r_c$. This choice highlights the region where photon trajectories can experience large deflections. Therefore, it is particularly useful for studying the brightness enhancement associated with the lensed image and the photon ring contributions. This region is also justified by the possible existence of orbits of massive particles between $r_I$ and $r_c$, specifically unstable orbits.
The third profile is peaked near the innermost stable circular orbit, $\mu_3\simeq r_{\rm ISCO}$. This profile represents a more standard thin disk configuration, in which most of the emission comes from the region where stable circular motion ceases. In this case, the emission is less concentrated near the horizon, and the central brightness depression tends to be more pronounced.

For each profile, the width $\sigma$ and the asymmetry parameter $\gamma$ can be chosen so that the emission is sufficiently localized around the desired radial scale while still allowing for smooth decay away from the peak. Since our main goal is to compare how the cloud of strings parameters modify the optical appearance, we keep the same functional form of $I^{em}(r)$ for all configurations and vary only the characteristic radii $r_h$, $r_c$, and $r_{I}$ according to the chosen spacetime parameters.

\begin{figure}[!htb]
    \centering
    \includegraphics[width=1\linewidth]{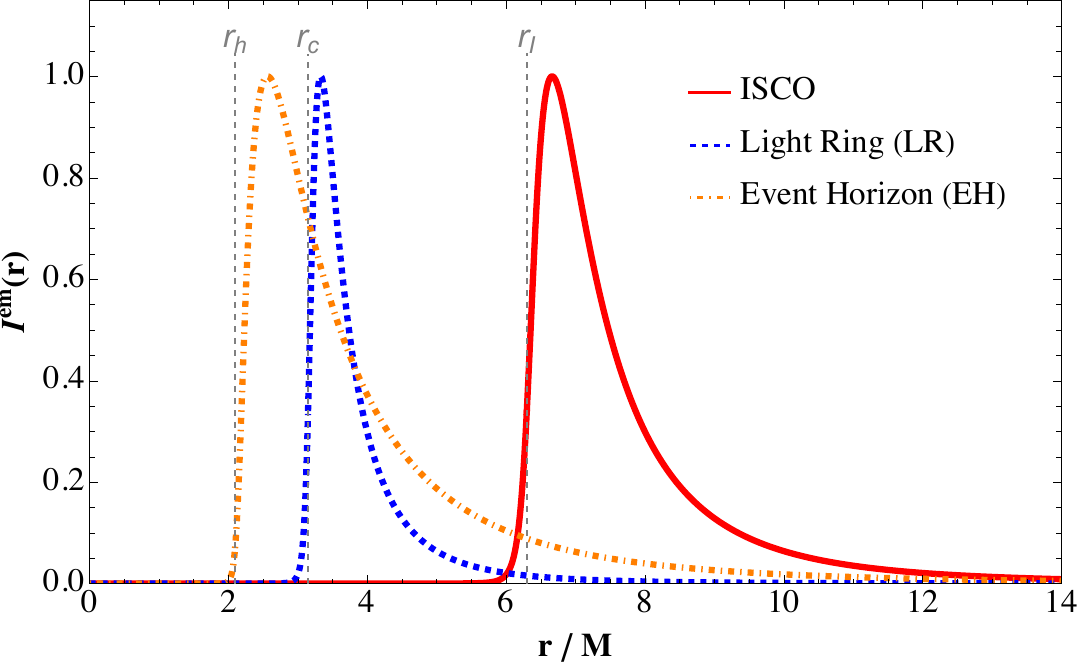}
    \caption{Behavior of the intensity profiles according to Eq.~\eqref{perfis}. For the profile parameters, we choose $\gamma=-1/2$ and $\sigma=M/4$. The cloud of strings parameters are taken as $l_s/M=1$ and $g_s=0.5$. For this choice of parameters, we have $r_h/M\simeq 2.096$, $r_c/M\simeq 3.147$, and $r_I/M\simeq 6.298$.
}
    \label{fig:perfis}
\end{figure}

The intrinsic luminosity profile $I^{em}(r)$ describes the radiation emitted locally by the disk. However, the observed intensity is not simply equal to $I^{em}(r)$. It must also include gravitational redshift and lensing effects. These effects will be incorporated later through the transfer functions $r_n(b)$, which determine the radial positions at which photons reaching the observer intersect the disk.

The first effect comes from the change in photon frequency between the emission point and the observer. According to Liouville's theorem, the quantity $I_\nu/\nu^3$ is conserved along null geodesics, where $I_\nu$ is the specific intensity, defined as the intensity per unit frequency, and $\nu$ is the photon frequency measured by a given observer. Therefore, the observed and emitted specific intensities are related by \begin{equation}
I_{\nu_{obs}}^{obs}= g^3 I_{\nu_{em}}^{em},
\end{equation}
where $g=\nu_{obs}/\nu_{em}$ is the redshift factor. If one considers the integrated frequency intensity, the observed intensity acquires an additional factor of $g$, so that $I^{obs}=g^4 I^{em}$.

For a static emitter and a static observer in the spacetime \eqref{line}, the redshift factor is determined only by the metric function $f(r)$. In this case, $g(r)=\sqrt{f(r)/f(r_O)}$,
where $r_O$ denotes the radial position of the observer. Hence, the intensity emitted at radius $r$ contributes to the observed intensity with the factor
\begin{equation}
g^4(r)=\left(\frac{f(r)}{f(r_O)}\right)^2.
\end{equation}
Thus, if a photon emitted at radius $r$ reaches the observer, its observed intensity is
\begin{equation}
I^{obs}(r)=\left[\frac{f(r)}{f(r_O)}\right]^2
I^{em}(r).
\end{equation}

In the Letelier--Alencar spacetime, the asymptotic value of the metric function is not unity, but instead
$f(r\to\infty)=1-g_s^2$. Therefore, for a distant observer, we have $f(r_O)\simeq 1-g_s^2$.

The observed intensity associated with emission at radius $r$ can then be rewritten as
\begin{equation}
I^{obs}(r)
\simeq
\left[
\frac{f(r)}{1-g_s^2}
\right]^2
I^{em}(r).
\end{equation}
This expression shows that the non-asymptotically flat character of the geometry modifies the normalization of the observed intensity. In the Schwarzschild limit, $g_s\to 0$, we recover $f(r_O)\to 1$, and the usual expression $I^{obs}=f(r)^2 I^{em}$ is obtained for a distant observer.

The second effect is due to the strong bending of photon trajectories. Since we consider an optically thin disk, photons are allowed to cross the equatorial plane more than once before reaching the observer \footnote{In the literature, there are works in which the authors compare the differences in the black hole appearance for optically thin and optically thick accretion disks \cite{Chen:2025wut}. Despite the clear differences between these disk models, in this work we focus only on the optically thin disk case.
}. Each crossing corresponds to a possible contribution to the observed intensity. Therefore, for a given impact parameter $b$, the total observed intensity is obtained by summing all intersections between the photon trajectory and the disk.

We denote by $r_n(b)$ the radial position of the intersection $n$-th of the photon trajectory with the accretion disk. The functions $r_n(b)$ are called transfer functions, since they map the position on the observer's screen, labeled by $b$, to the emission radius on the disk. The first intersection corresponds to the direct emission, the second one to the lensed emission, and higher order intersections are associated with photon ring contributions.

Therefore, the total observed intensity can be written as
\begin{equation}
I^{obs}(b)=\sum_n \left[
\frac{f(r_n(b))}{f(r_O)} \right]^2 I^{em}(r_n(b)).
\end{equation}
For a distant observer in the Letelier--Alencar spacetime, this becomes
\begin{equation}
I^{obs}(b)
\simeq
\sum_{n=0}^{N}
\left[
\frac{f(r_n(b))}{1-g_s^2}
\right]^2
I^{em}(r_n(b)),
\end{equation}
where $N$ is chosen according to the desired numerical accuracy. In this work, we truncate the sum at $N=2$. This choice includes the direct emission, the first lensed contribution, and the first photon ring contribution, while neglecting higher order crossings. Such higher order terms correspond to photons that perform additional half orbits around the black hole before reaching the observer. They are increasingly concentrated near the critical impact parameter and their contribution to the total luminosity becomes progressively smaller. Therefore, keeping terms up to $N=2$ is sufficient to capture the dominant optical features of the image while avoiding unnecessary numerical complications. The term $n=0$ represents the direct image of the disk, $n=1$ represents the first lensed contribution, and $n=2$ represents the first photon ring contribution.

The transfer functions will be written in terms of the impact parameter $b$, since this quantity naturally labels the null geodesics reaching the observer. For a fixed observer position, this is equivalent to using the radial coordinate on the observer's screen, up to a constant rescaling. Indeed, for a static observer located at $r_O$, the screen radius is given, in the small-angle approximation, by
\begin{equation}
R_{\rm obs}=b\sqrt{f(r_O)}=b\sqrt{1-g_s^2}.
\end{equation}
Thus, using $b$ as the argument of the transfer functions is convenient for the ray-tracing procedure, while the corresponding physical screen coordinate can be obtained by this simple rescaling. This is the same reasoning adopted at the end of the section on the black hole shadow radius, where we discussed why it is not equal to the critical impact parameter.

\begin{figure}[!htb]
    \centering
    \includegraphics[width=0.95\linewidth]{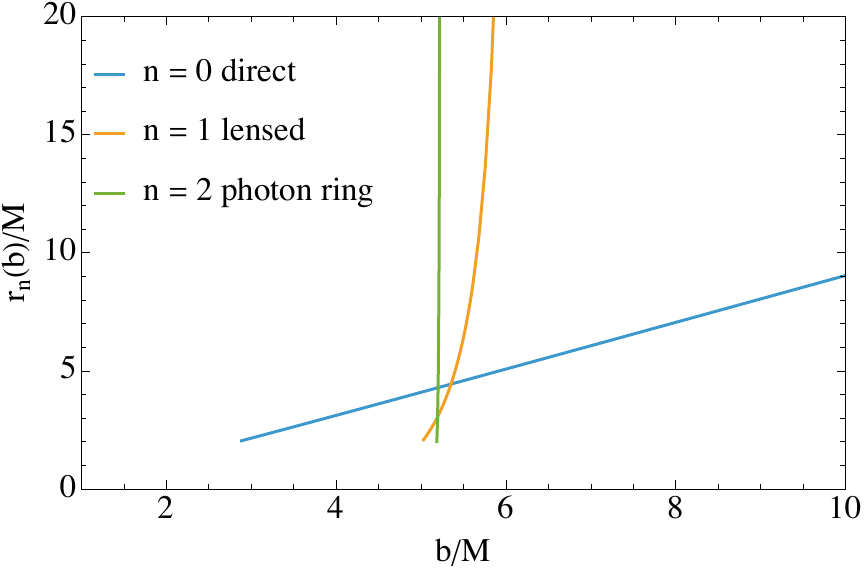}
    \includegraphics[width=0.95\linewidth]{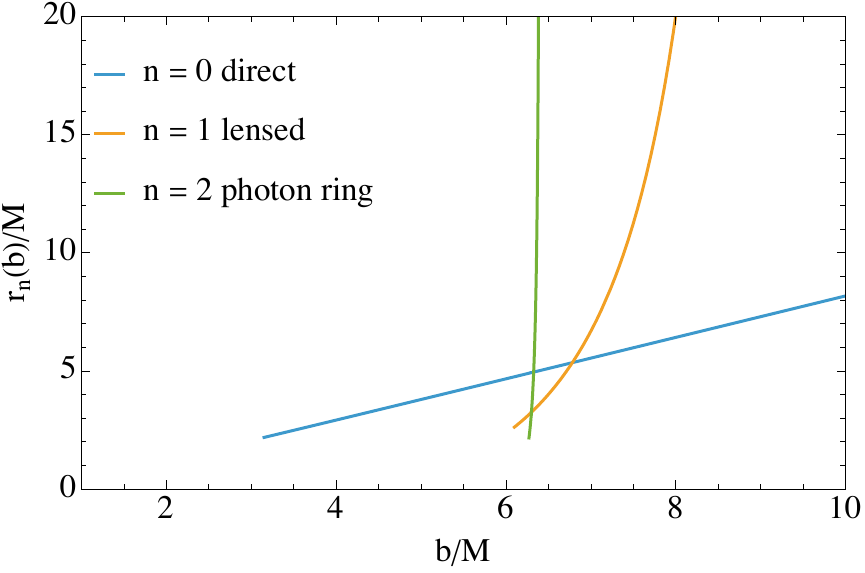}
    \includegraphics[width=0.95\linewidth]{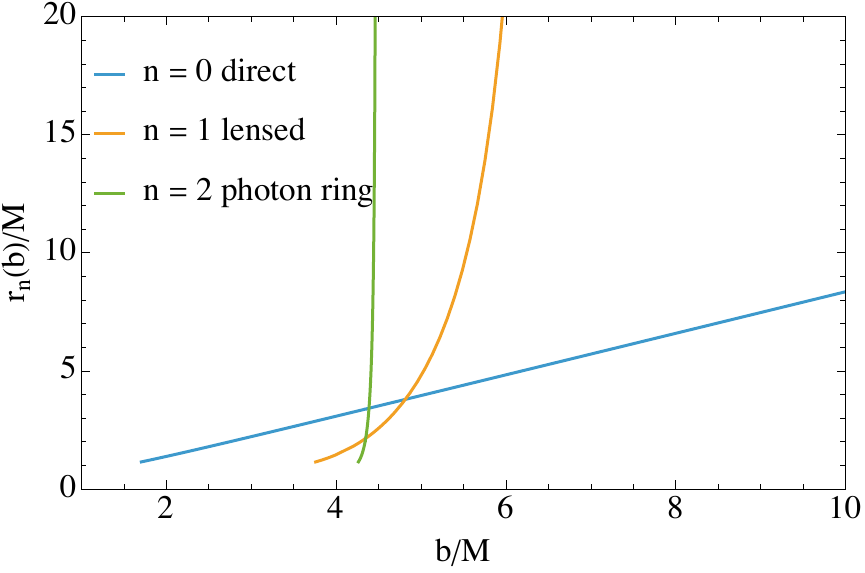}
    \caption{Behavior of the transfer function as a function of the impact parameter. In the top panel, we consider the Schwarzschild case. In the middle panel, we choose $l_s/M=1$ and $g_s=0.5$. In the bottom panel, we fix $l_s/M=2$ and $g_s=0.5$.}
    \label{fig:transfer}
\end{figure}

The transfer functions shown in Fig.~\ref{fig:transfer} reveal that the generalized cloud of strings modifies the mapping between the emitting regions of the disk and the observer's screen in a nonuniform manner for different image orders. In the Schwarzschild case, the direct contribution extends over a broad range of impact parameters, whereas the lensed and photon ring contributions are strongly concentrated near the critical region. For $l_s=M$ and $g_s=0.5$, the transfer functions are shifted toward larger impact parameters, and the lensed contribution extends over a wider interval, suggesting a broader redistribution of strongly deflected radiation on the observer's screen. By contrast, for $l_s=2M$ and $g_s=0.5$, both the contributions move toward smaller impact parameters, accompanied by a more noticeable modification of the direct branch. Therefore, the cloud of strings parameters do not simply produce an overall rescaling of the optical image, but rather alter the relative mapping of the direct, lensed, and photon ring contributions. These changes are expected to affect the location and width of the bright structures in the observed intensity profile, with the precise pattern also depending on the adopted emission model and gravitational redshift.

\begin{figure*}[!htb]
    \includegraphics[width=0.5\linewidth]{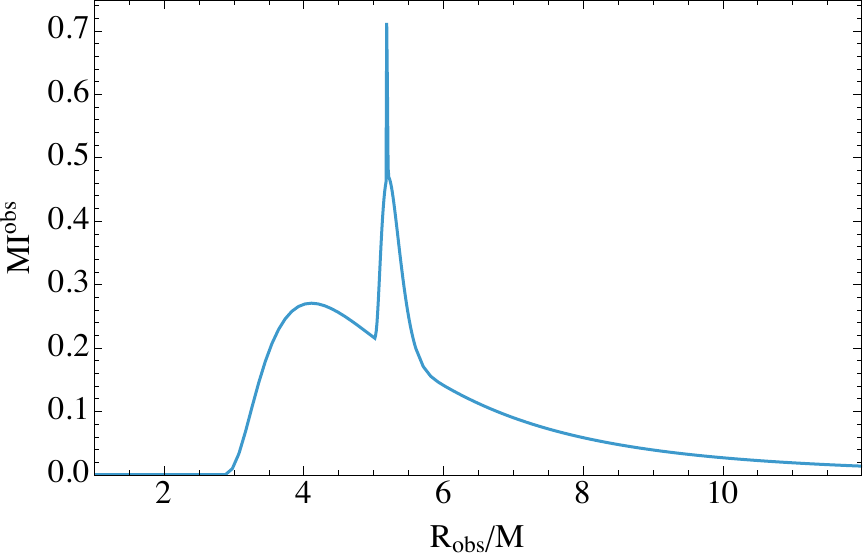}
    \includegraphics[width=0.4\linewidth]{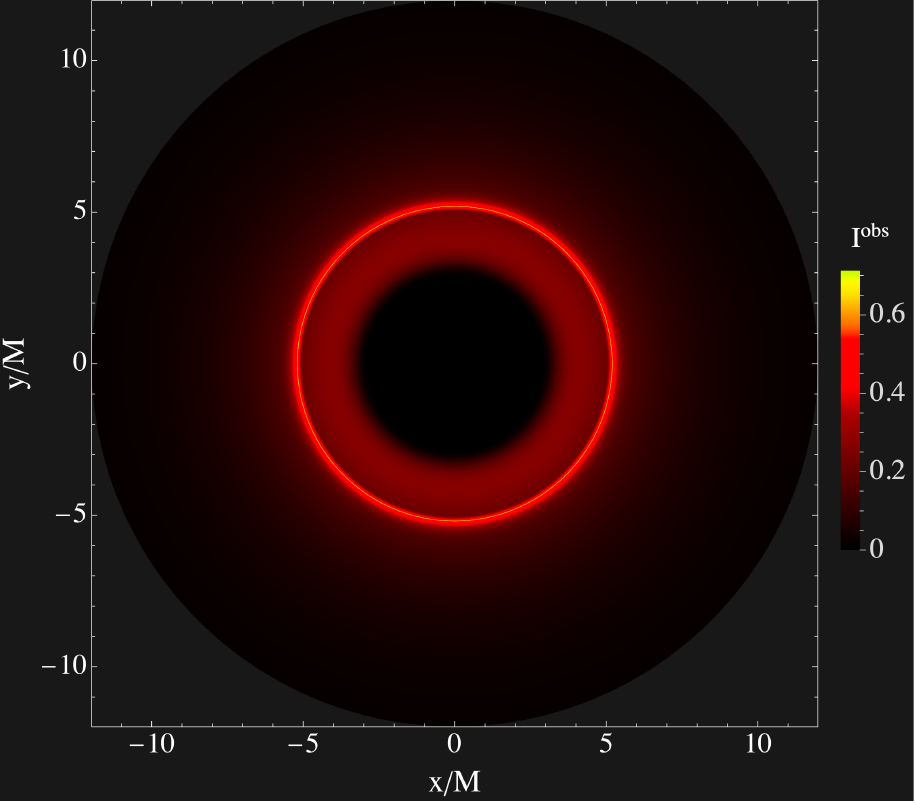}
    \includegraphics[width=0.5\linewidth]{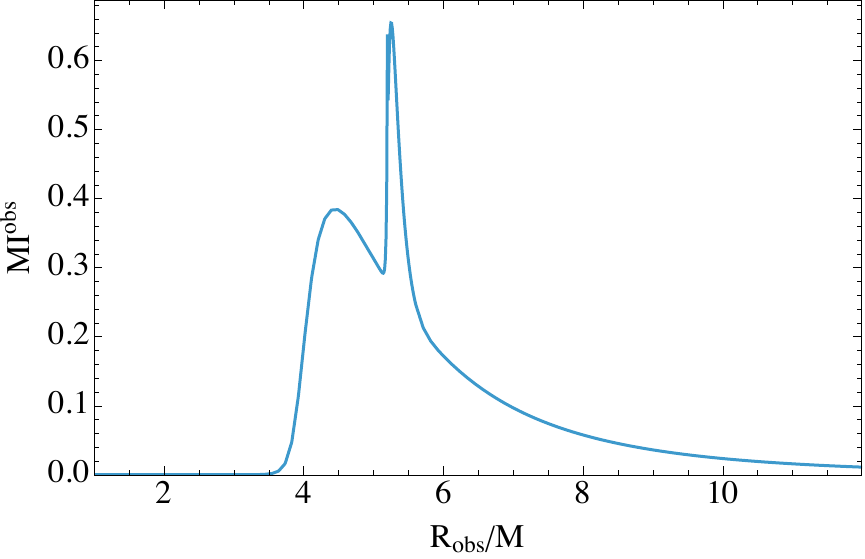}
    \includegraphics[width=0.4\linewidth]{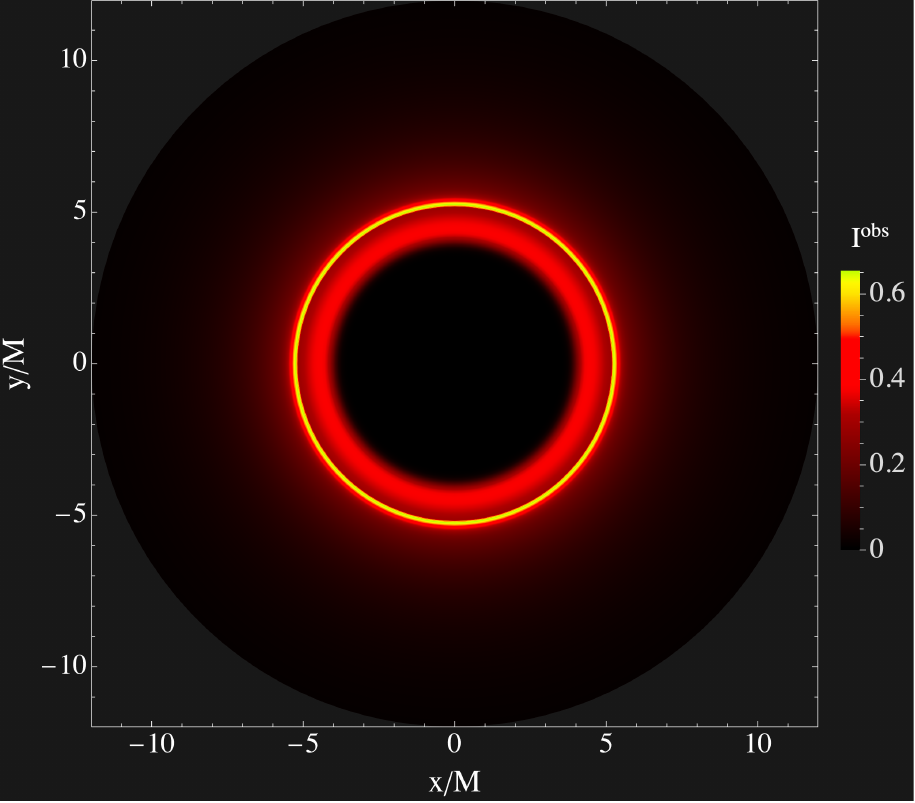}
    \includegraphics[width=0.5\linewidth]{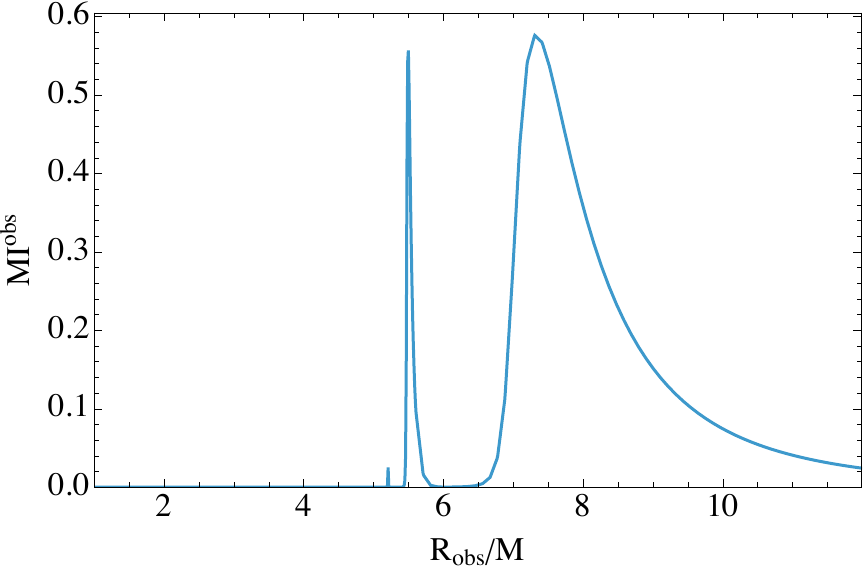}
    \includegraphics[width=0.4\linewidth]{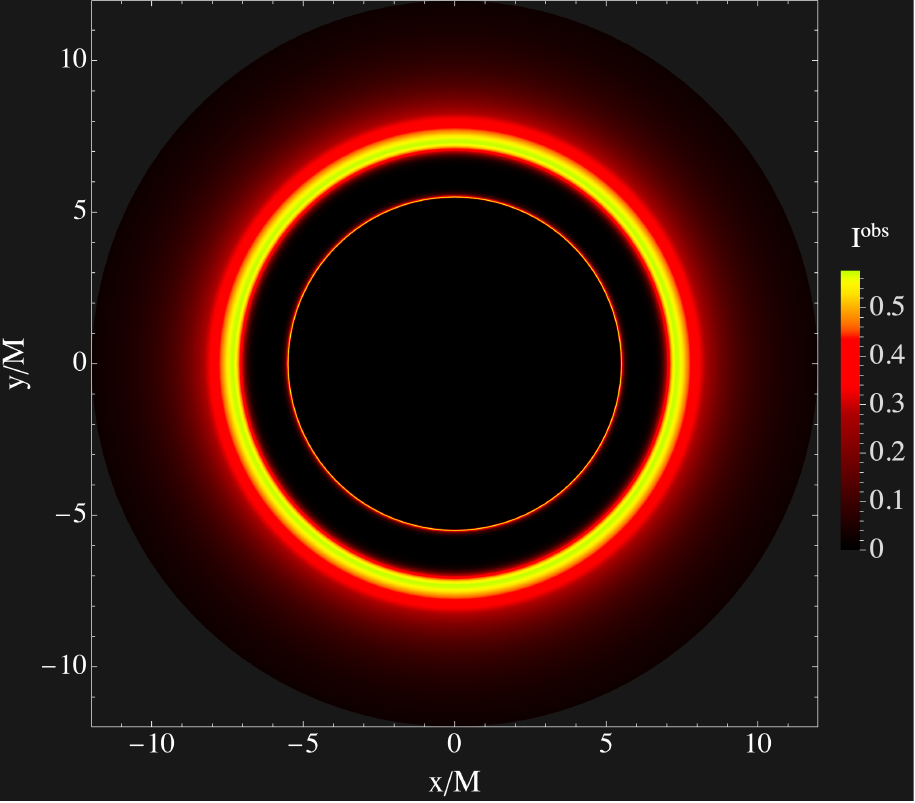}
    \caption{Observed intensity profiles (left panels) and optical appearances (right panels) for the Schwarzschild case. From top to bottom, we consider, respectively, the profiles whose luminosity peaks are located at $r=r_h=2M$, $r=r_c=3M$, and $r=r_I=6M$.
}
    \label{fig:shadow_Schwarzschild}
\end{figure*}

\begin{figure*}[!htb]
    \includegraphics[width=0.5\linewidth]{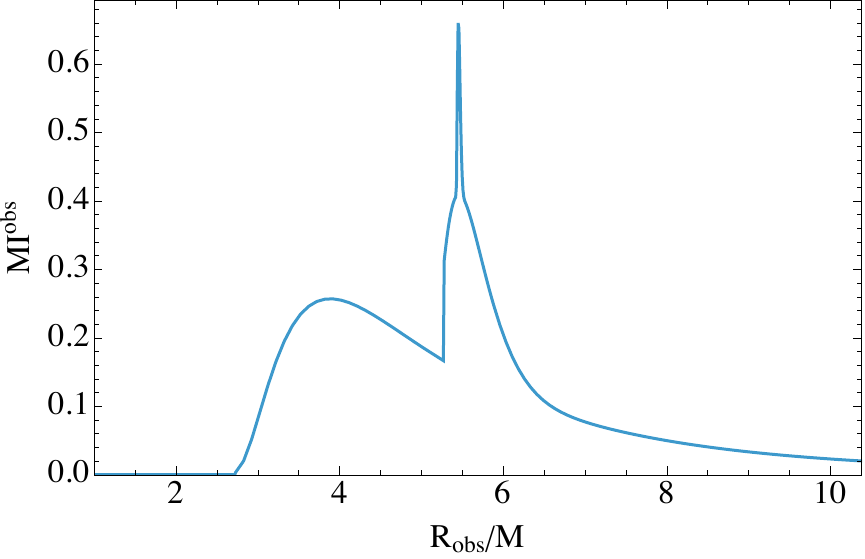}
    \includegraphics[width=0.4\linewidth]{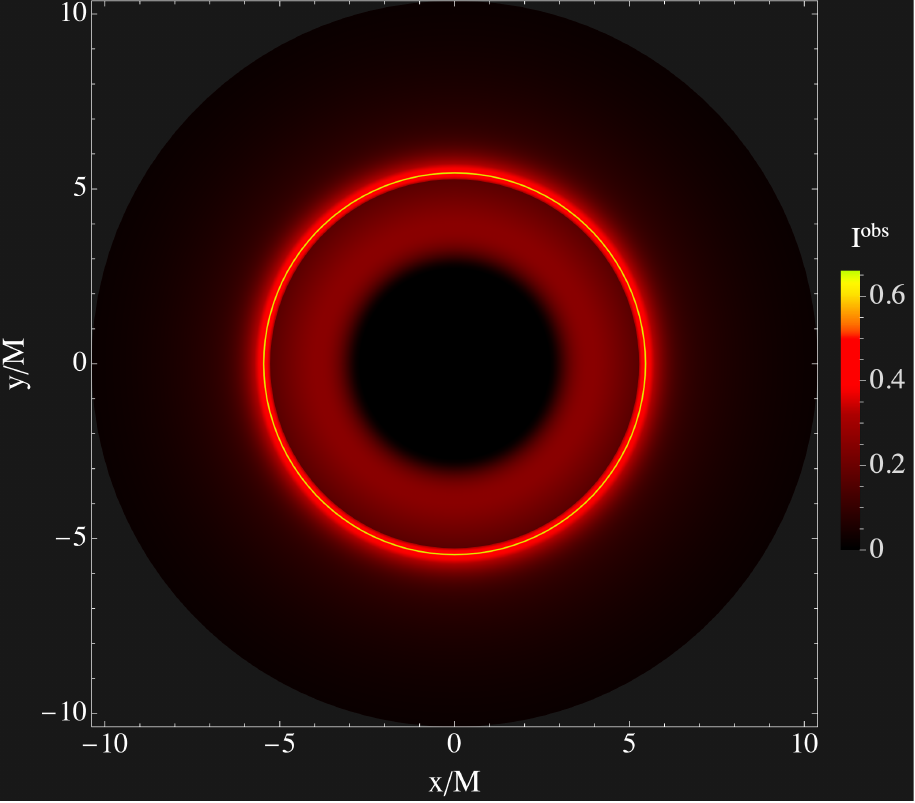}
    \includegraphics[width=0.5\linewidth]{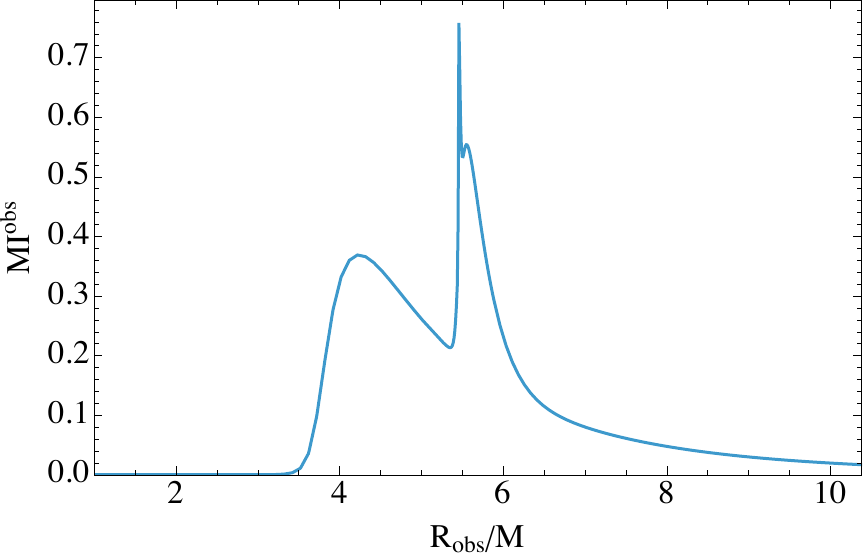}
    \includegraphics[width=0.4\linewidth]{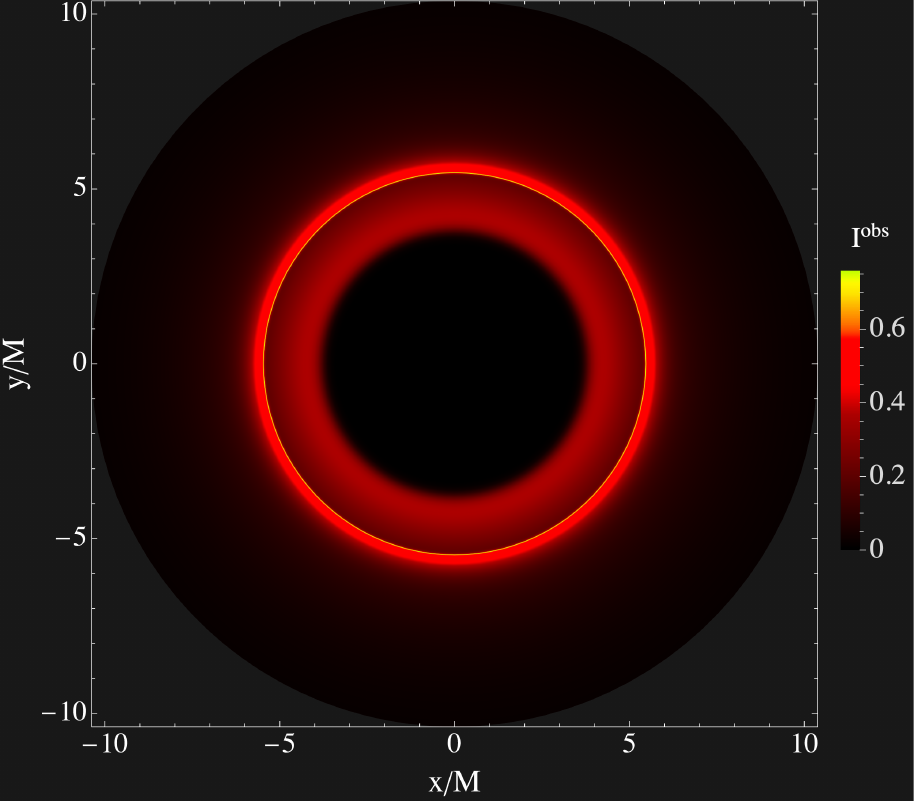}
    \includegraphics[width=0.5\linewidth]{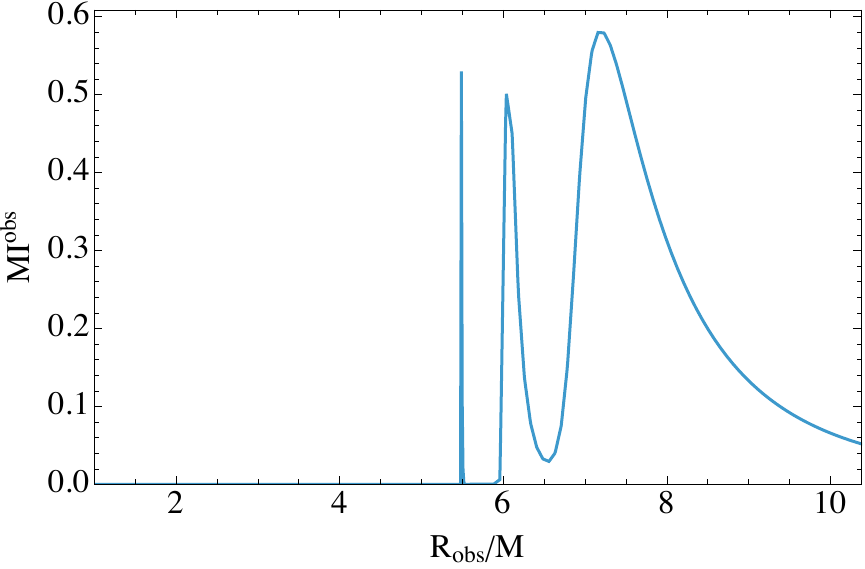}
    \includegraphics[width=0.4\linewidth]{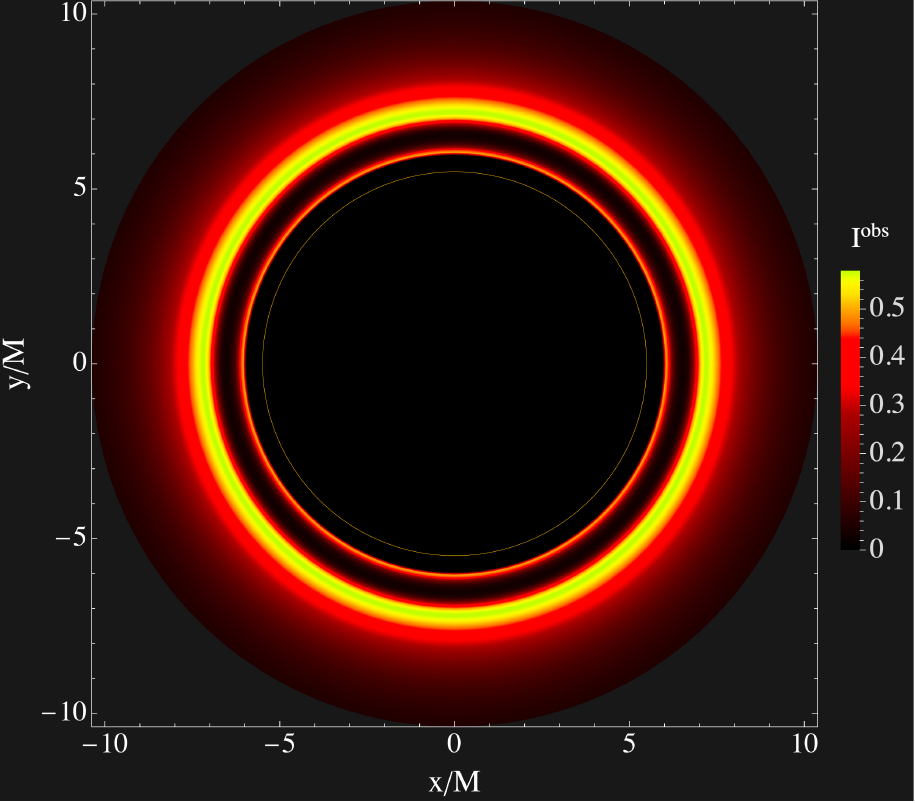}
    \caption{Observed intensity profiles (left panels) and optical appearances (right panels) for the Letelier--Alencar case with $l_s=M$ and $g_s=0.5$. From top to bottom, we consider, respectively, the profile whose luminosity peak is located at $r=r_h\simeq 2.096M$, $r=r_c\simeq 3.147M$, and $r=r_I\simeq 6.298M$.
}
    \label{fig:shadow_ls1}
\end{figure*}

\begin{figure*}[!htb]
    \includegraphics[width=0.5\linewidth]{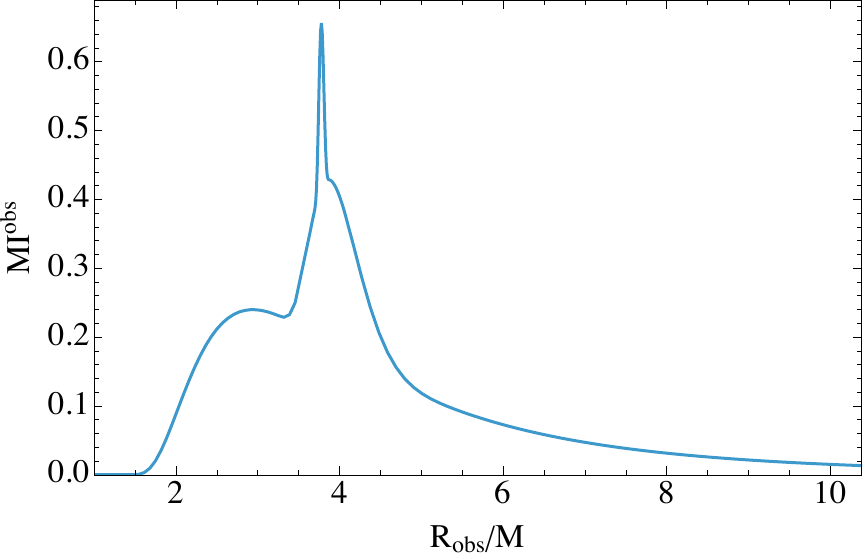}
    \includegraphics[width=0.4\linewidth]{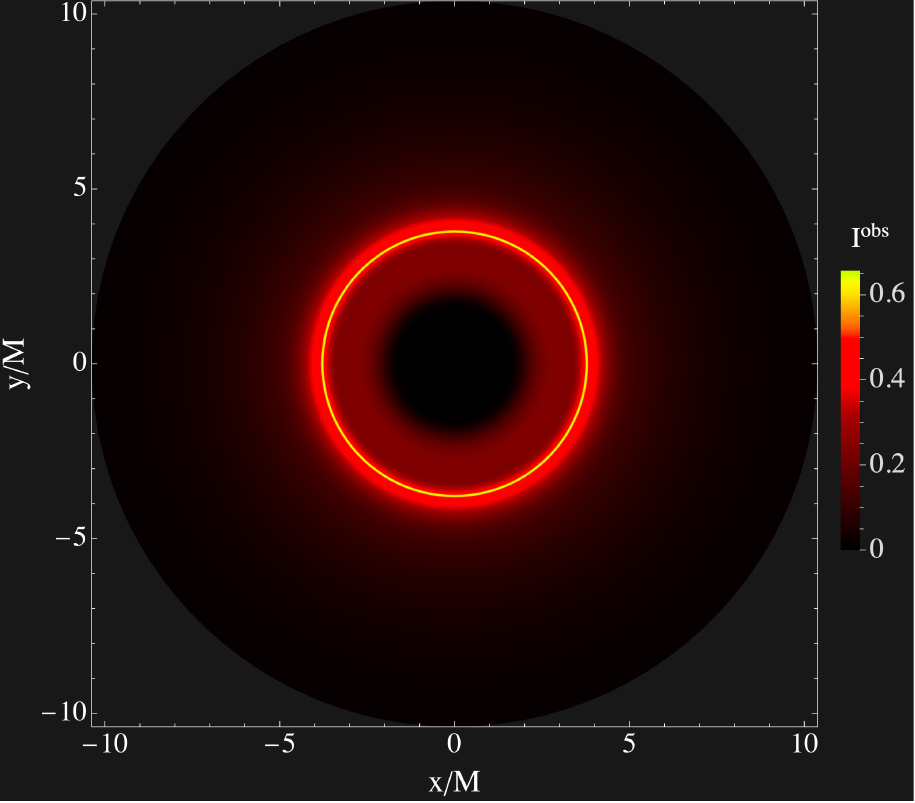}
    \includegraphics[width=0.5\linewidth]{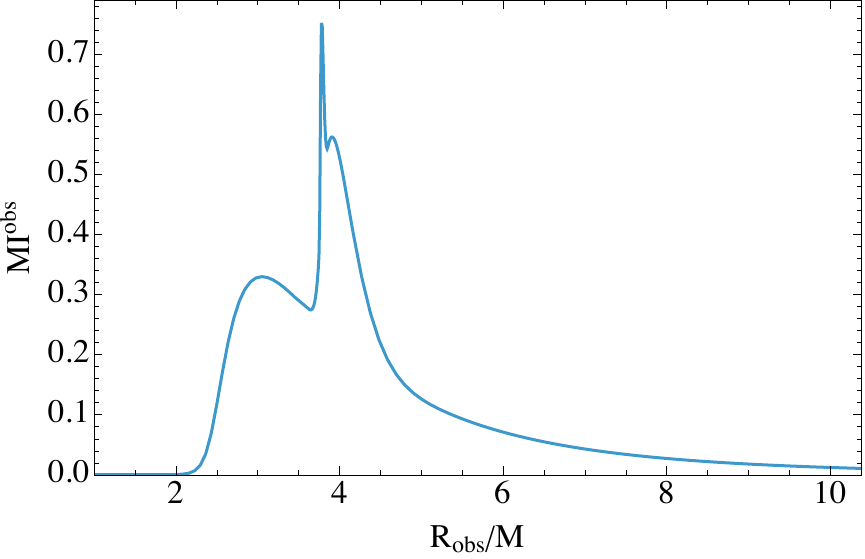}
    \includegraphics[width=0.4\linewidth]{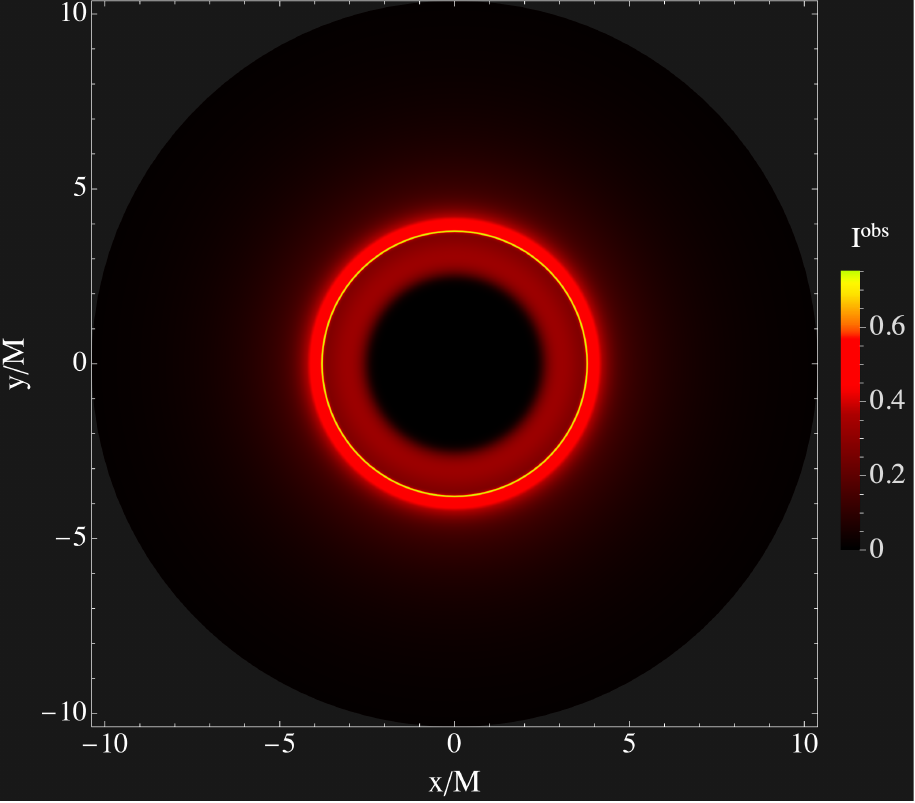}
    \includegraphics[width=0.5\linewidth]{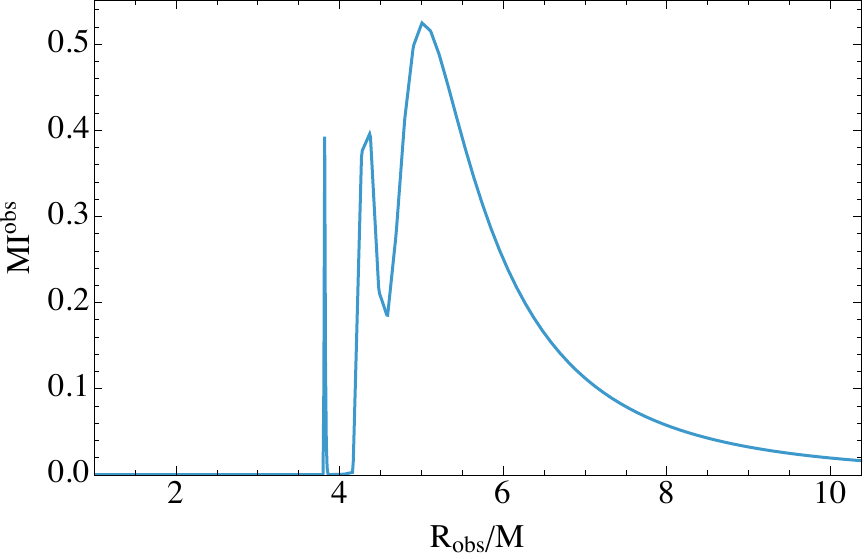}
    \includegraphics[width=0.4\linewidth]{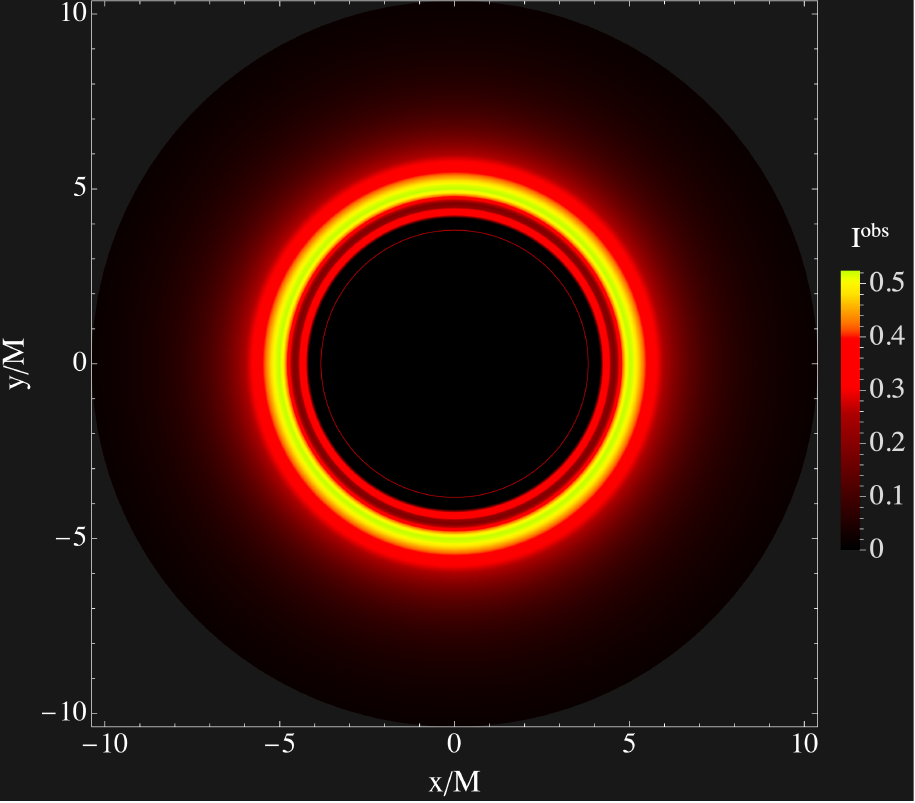}
    \caption{Observed intensity profiles (left panels) and optical appearances (right panels) for the Letelier--Alencar case with $l_s=2M$ and $g_s=0.5$. From top to bottom, we consider, respectively, the profile whose luminosity peak is located at $r=r_h\simeq 1.129M$, $r=r_c\simeq 2.023M$, and $r=r_I\simeq 4.262M$.}
    \label{fig:shadow_ls2}
\end{figure*}

The observed intensity profiles and the corresponding optical appearances are shown in Fig.~\ref{fig:shadow_Schwarzschild}, for the Schwarzschild case, Fig.~\ref{fig:shadow_ls1}, for the Letelier--Alencar black hole with $ls=M$ and $g_s=0.5$, and Fig.~\ref{fig:shadow_ls2}, for the Letelier--Alencar black hole with $l_s=2M$ and $g_s=0.5$, with the three emission profiles considered before. In the Schwarzschild spacetime, emission concentrated near the event horizon or the photon sphere produces a broad direct component accompanied by a narrow peak close to the critical screen radius. The latter originates from higher order images, whose transfer functions map a comparatively wide range of emission radii into a very narrow interval on the observer's screen. When the emission is concentrated near the ISCO, the direct contribution is instead mapped to larger screen radii, producing a broad outer bright ring, while the lensed and photon ring contributions remain concentrated near the critical region and generate additional narrow inner rings. The presence of the generalized cloud of strings significantly modifies this radial mapping. For $l_s=M$ and $g_s=0.5$, the critical structures are slightly shifted outwards, and the lensed contribution is distributed over a broader radial interval. In contrast, for $l_s=2M$ and $g_s=0.5$, both the direct and higher order structures are displaced toward considerably smaller screen radii, resulting in a more compact optical appearance. These results show that the cloud of strings parameters affect not only the characteristic size of the central dark region, but also the relative separation and radial distribution of the direct, lensed, and photon ring images. The detailed brightness pattern remains sensitive to the assumed emission profile, whereas the displacement of the higher order structures directly reflects the modification of the null geodesic mapping induced by the spacetime geometry. It is also possible to note that for the ISCO profile, there is an innermost luminosity peak in cases with a cloud of strings.

\subsection{Gravitational lensing and celestial sphere images}

The optical appearance discussed above depends on both the spacetime geometry and the emission model adopted for the accretion disk. Although this approach provides information on the observable intensity distribution, the detailed brightness pattern remains sensitive to the assumed properties of the emitting matter. To isolate more directly the effects of the background geometry on photon propagation, we now investigate the gravitational lensing of a celestial sphere surrounding the black hole.

The images are constructed through a standard backward ray-tracing procedure \cite{Cunha:2016bpi,Cunha:2015yba,Cunha:2018acu}. In this setup, a static observer is placed at a finite radial position $r_{\rm O}$, while a spherical distribution of light sources is located at $r_{\rm CS}>r_{\rm O}$ \cite{Lima:2021las}. Each direction in the observer's local sky, defined with respect to the orthonormal frame of the static observer, determines the initial direction of a null geodesic, which is integrated backwards from the observer \cite{dePaula:2023ozi,dePaula:2026blu}. Rays that reach the celestial sphere are assigned the color of their corresponding intersection point, whereas trajectories captured by the black hole form the shadow region. The resulting image therefore provides a representation of the lens map between the observer's local sky and the celestial sphere.

To facilitate the visualization of the lensing structure, the celestial sphere is divided into four colored quadrants and covered by a regular angular grid. A bright extended spot is placed along the direction opposite to the observer, such that the source, the black hole, and the observer are approximately aligned. This reference source provides a way of identifying strongly lensed structures and multiple images. We adopt the same observational setup for the Schwarzschild and Letelier--Alencar geometries, fixing the observer at $r_{\rm O}=15M$. Therefore, any differences between the resulting images arise from the modification of photon propagation induced by the cloud of strings parameters.

\begin{figure}[!htb]
    \centering
    \includegraphics[width=0.79\linewidth]{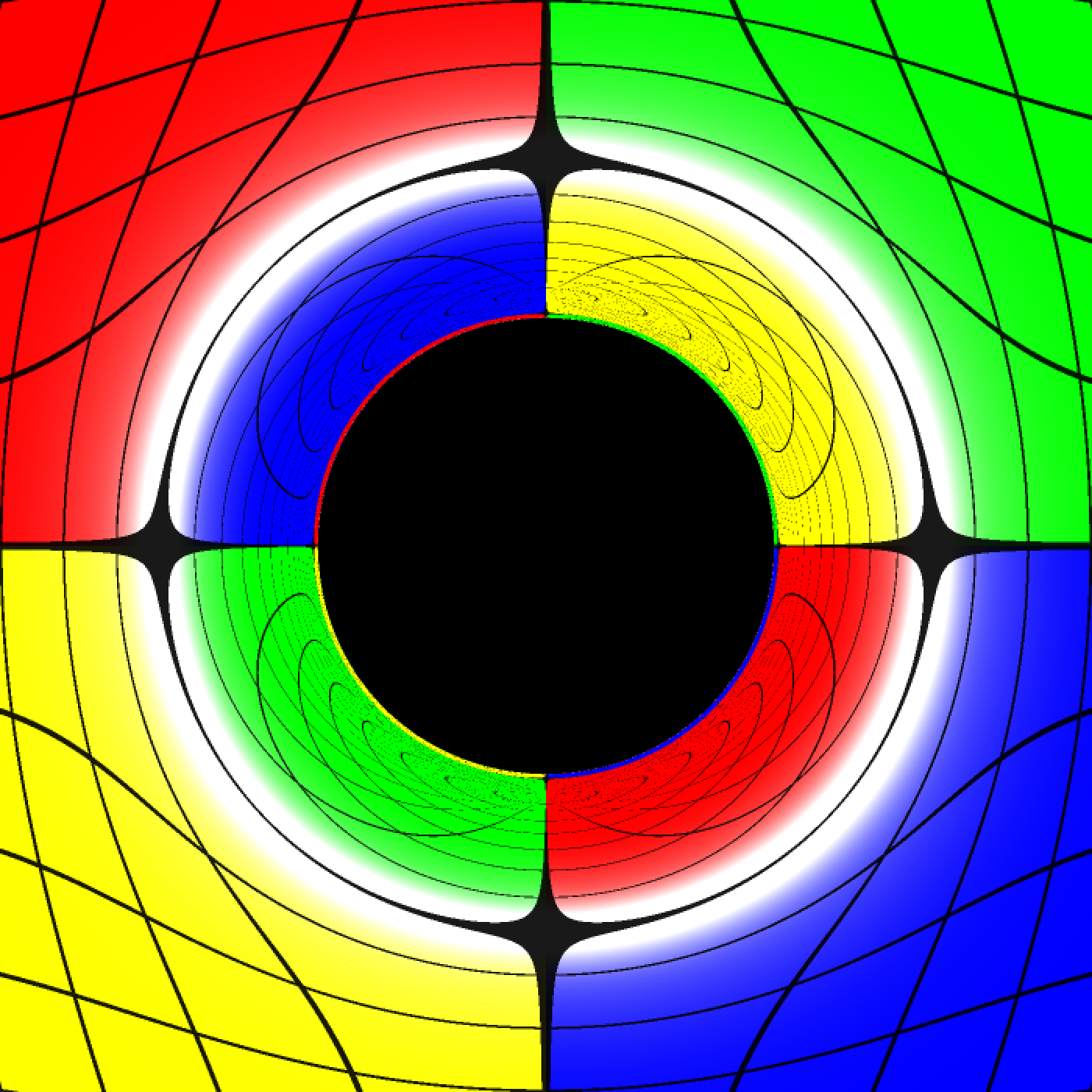}
    \includegraphics[width=0.79\linewidth]{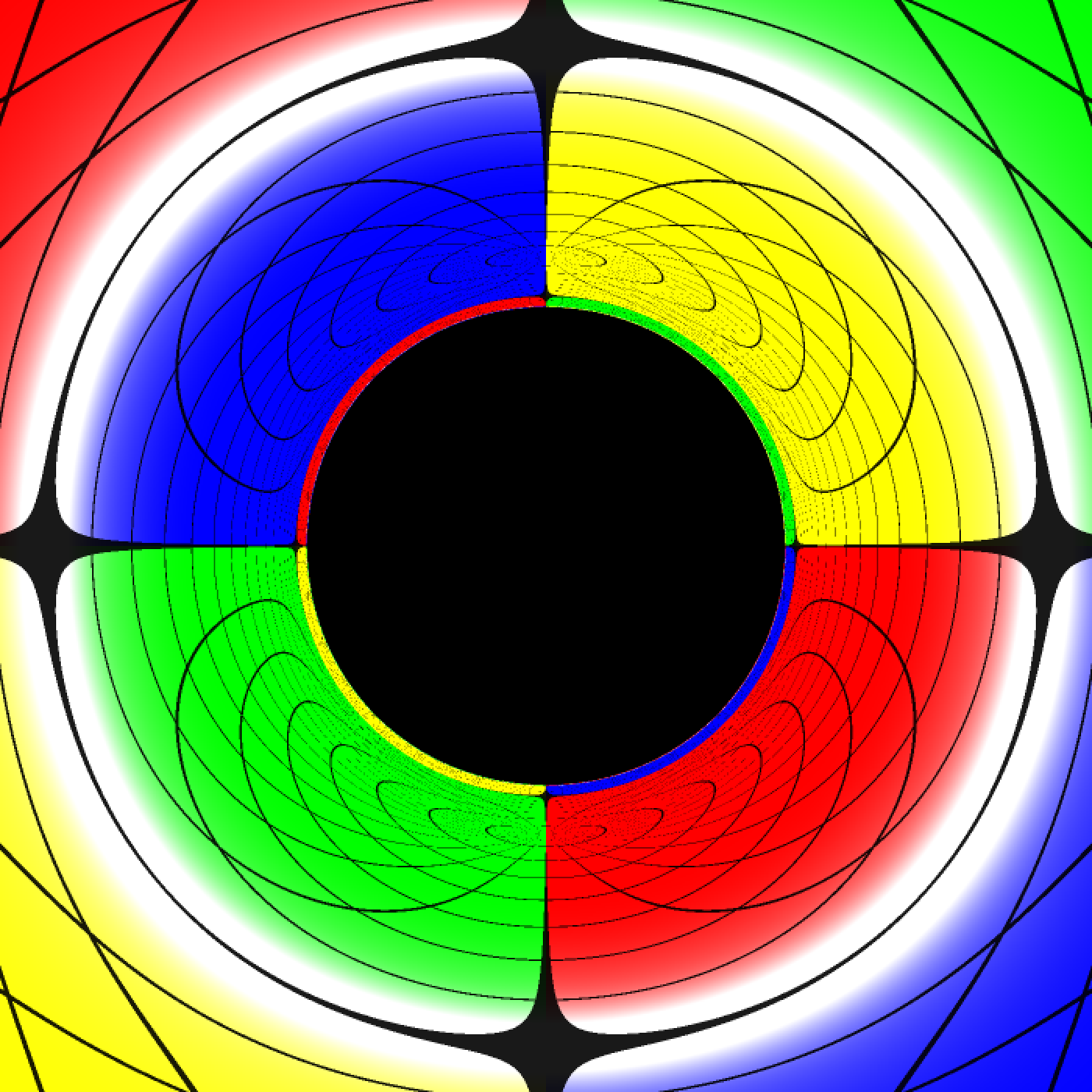}
    \includegraphics[width=0.79\linewidth]{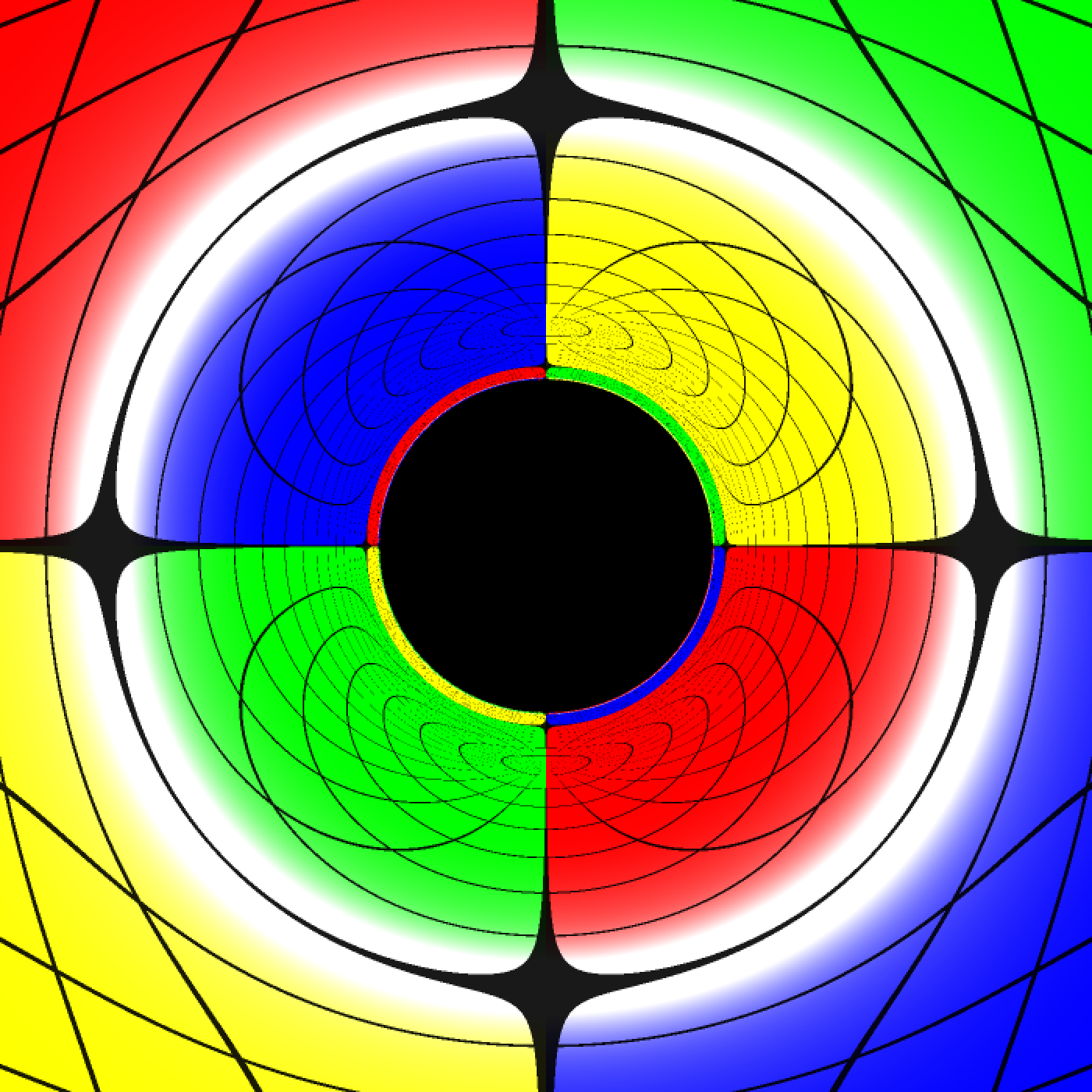}
    \caption{Celestial sphere images obtained through backward ray tracing. From top to bottom: Schwarzschild case, Letelier--Alencar black hole with $l_s=M$ and $g_s=0.5$, and Letelier--Alencar black hole with $l_s=2M$ and $g_s=0.5$. In all cases, the observer is located at $r_{\rm O}=15M$.}
    \label{fig:lensing}
\end{figure}

The resulting images are shown in Fig.~\ref{fig:lensing}. The central dark region corresponds to the black hole shadow, the set of directions on the observer's local sky associated with null geodesics that are captured by the black hole. Therefore, the boundary of this region should not be interpreted as a direct image of the event horizon or of the photon sphere. Instead, it represents the critical curve separating captured trajectories from those that escape and reach the celestial sphere. 

As expected from spherical symmetry, the shadow remains circular in all three configurations. Nevertheless, its angular size is modified by the generalized cloud of strings. For $l_s=M$ and $g_s=0.5$, the shadow is larger than in the Schwarzschild case, consistently with the outward displacement of the event horizon, photon sphere, and critical null geodesic scale found in the previous sections. On the other hand, for $l_s=2M$ and $g_s=0.5$, the central dark region is considerably smaller, in agreement with the reduction of the corresponding critical scales.

The broad white annular structure surrounding the shadow is produced by the gravitationally lensed image of the bright reference source placed on the celestial sphere approximately behind the black hole. Due to the nearly axial alignment between the source, the lens, and the observer, spherical symmetry maps the source into an Einstein ring. Since the reference source has a finite angular extension rather than being pointlike, its image does not form an infinitesimally thin Einstein ring, but instead appears as a broadened luminous annulus. Finer structures accumulating near the shadow edge are associated with more strongly deflected trajectories and higher order images.

An interesting feature of Fig.~\ref{fig:lensing} is that the angular position of this bright annulus does not follow the same trend as the shadow size. For $l_s=M$ and $g_s=0.5$, the Einstein ring structure is shifted outward with respect to the Schwarzschild case, which is qualitatively consistent with the increase of the characteristic critical scales of the geometry. However, for $l_s=2M$ and $g_s=0.5$, the shadow becomes smaller than the Schwarzschild shadow while the main lensed image of the bright source still appears at a larger angular distance from the center.

This apparently opposite behavior reflects the fact that the shadow and the Einstein ring image probe different aspects of null geodesic propagation. The shadow boundary is controlled by the critical null orbit separating capture from escape and is therefore directly related to the critical impact parameter. In contrast, the angular position of the lensed image of the reference source is determined by the complete lens map connecting the observer to the celestial sphere. It depends on the accumulated deflection along the entire photon trajectory and not solely on the properties of the unstable circular null orbit. Consequently, a decrease in the shadow size does not necessarily imply a corresponding inward displacement of the lensed images.

The deformation of the angular grid further illustrates this nonuniform modification of the lens map. Far from the central region, the three configurations exhibit qualitatively similar patterns, whereas the differences become increasingly pronounced in the strong-field region surrounding the shadow. The cloud of strings parameters modify both the angular compression of the background grid and the position of the strongly lensed images. These results are consistent with the behavior previously observed in the transfer functions, where the generalized cloud of strings was shown to modify the mapping between emission radii and the observer's screen in a nonuniform manner. The celestial sphere images therefore demonstrate that the optical effects of the generalized cloud of strings cannot be described as a simple overall rescaling of the Schwarzschild lensing pattern.
\section{Summary and conclusion}\label{Sec:Conclusion}

In this work, we investigated the optical properties of the Letelier--Alencar black hole, which describes a static and spherically symmetric black hole surrounded by a generalized cloud of strings. The cloud distribution is characterized by the effective coupling $g_s$ and the length scale $l_s$, whose combined effects modify photon propagation both in the vicinity of the black hole and in the asymptotic region. Since the metric does not behave as the Minkowski geometry at spatial infinity, particular care is required when interpreting quantities.

We first analyzed the null geodesic structure through the effective potential, the unstable circular photon orbit, and the critical impact parameter. The effects of the two cloud of strings parameters were found to compete with each other. For sufficiently small values of $l_s$, increasing $g_s$ generally increases the
photon sphere radius and the critical impact parameter. However, as $l_s$ becomes larger, this behavior is progressively suppressed and may eventually be reversed, so that increasing $g_s$ leads to smaller characteristic photon orbit scales. Similarly, for fixed $g_s$, increasing $l_s$ tends to reduce both the photon sphere radius and the critical impact parameter. The photon trajectories are consistent with the effective potential analysis, showing that the parameters of the cloud of strings modify the boundary between captured and scattered trajectories.

We then determined the shadow radius for a static observer located at a finite radial distance. An important result is that the critical impact parameter cannot be directly identified with the physical shadow radius in this geometry. This distinction arises from the non Minkowskian asymptotic behavior, $f(r)\rightarrow 1-g_s^2$, which introduces an additional factor in the conversion between the impact parameter and the observer's screen coordinate. By comparing the theoretical shadow radius with the observational intervals inferred for Sgr~A* and M87*, we obtained constraints on the parameter space of the model. The Sgr A* bounds provide the more restrictive test because of their narrower observational intervals. A non negligible region of the $(g_s,l_s)$ parameter space remains compatible with the current shadow measurements.

In the weak-field regime, we calculated the deflection angle using two complementary approaches: the Gauss--Bonnet method applied to the optical geometry and a direct perturbative expansion of the null geodesic equations. The two calculations yield the same result at the order considered, providing an independent consistency check. The deflection angle contains a contribution associated with the asymptotically conical optical geometry in addition to the local term generated by the curvature surrounding the black hole. The Schwarzschild and original Letelier results are recovered in the limits $g_s\rightarrow0$ and $l_s\rightarrow0$, respectively. These results emphasize that, in a non asymptotically flat spacetime, the total angular deflection cannot be characterized solely by the local curvature of the black hole region. We also investigated the Shapiro time delay. Within the weak-field and large distance approximations adopted
in our analysis, the presence of the cloud of strings introduces a cumulative contribution associated with the
non asymptotically flat background. In contrast with the standard Schwarzschild logarithmic contribution, this additional term grows with the propagation distance and is mainly controlled by $g_s$. For the configurations considered, the dependence on $l_s$ enters through a subleading correction and is therefore considerably weaker. The numerical estimates for Sgr~A* and M87* are consistent with this analytical behavior. These results indicate that precision measurements of light propagation times could provide a complementary probe of cloud of strings geometries.

Finally, we studied the observable appearance of the black hole using geometrically and optically thin accretion disks and backward ray tracing of a celestial sphere. The transfer functions show that the generalized cloud of strings modifies the mapping between the emission radius and the observer's screen differently for the direct, lensed, and photon ring contributions. The resulting images are therefore not obtained through a uniform rescaling of the corresponding Schwarzschild image. Depending on the values of $g_s$ and $l_s$, the positions, widths, and relative intensities of the observed features can be shifted in distinct ways.

The celestial sphere images further illustrate this nonuniform behavior. In particular, we found configurations for which the shadow is smaller than in Schwarzschild, while the main Einstein ring image of the extended background source is displaced outward. This is not contradictory, since the shadow boundary is governed by the critical null orbit separating
capture from escape, whereas the position of the lensed source image depends on the complete photon trajectory between the
source and the observer. Consequently, the shadow size, the positions of lensed images, and the intensity distribution probe different aspects of null geodesic propagation and may provide complementary information about the parameters of the generalized cloud of strings.

Overall, our results demonstrate that the optical signatures of the Letelier--Alencar geometry cannot be characterized by a single length scale or by a simple deformation of the Schwarzschild shadow. A joint analysis is therefore required to distinguish the effects of $g_s$ and $l_s$. Future investigations could extend the present analysis to rotating configurations and more realistic emission models, including inclined disks, plasma effects, and radiative-transfer calculations. Such extensions would make it possible to assess more accurately whether the signatures identified here can be constrained by present and future high resolution observations of black hole environments. It is also interesting to analyze the behavior of fields in the vicinity of the black hole through phenomena such as absorption, scattering, and quasinormal modes. These studies are also important for assessing the stability of the solution.

\section*{Acknowledgements}

M. S. and E. L. B. Junior would like to thank the Conselho Nacional de Desenvolvimento Cient\'ifico e Tecnol\'ogico - CNPq, Brazil for its financial support. M. S. was supported by CNPq/PDE: 200218/2025-5 and E. L. B. Junior by CNPq/PQ: 307085/2026-0. This work is also supported by the Spanish National Grant PID2024-157196NB-I00 funded by MICIU/AEI/10.13039/501100011033 and by the Department of Education, Junta de Castilla y Le\'on and FEDER Funds, Ref. CLU-2023-1-05.


\bibliography{ref.bib}

\end{document}